\documentclass{jfm}
\usepackage{graphicx}
\usepackage{epstopdf, epsfig}
\usepackage{hyperref}
\usepackage{bm}
\usepackage{color,soul}
\usepackage{mathrsfs}
\usepackage{amsmath,amssymb}

\shorttitle{Drop impact on heated substrate}

\shortauthor{D. Roy, S. Rao, V. Hariharan and S. Basu}

\title{Insights into air cushion dynamics during drop impact on heated substrate at low impact energy}

\author{Durbar Roy\aff{1},
Srinivas Rao S\aff{1},
Vishnu Hariharan \aff{1},
\and Saptarshi Basu\aff{1}
  \corresp{\email{sbasu@iisc.ac.in}}}

\affiliation{
\aff{1}Department of Mechanical Engineering, Indian Institute of Science, Bengaluru, 560012, India
}

\begin{document}

\maketitle

\begin{abstract}
We study the air layer dynamics beneath a low kinetic energy drop impinging a heated
substrate using high-speed reflection interferometry imaging and theoretical analysis. The
air layer has been subdivided into two distinct disjoint regions, the central dimple and its
immediate neighbourhood the peripheral disc. We decipher that a Gaussian profile can
approximate the dynamic shape evolution of the central air dimple. The dimple geometry
is a function of impact energy and its dependence on surface temperature is relatively
weak. The air layer rupture time and rupture radius increases with increase in substrate
temperature. We characterize the air layer profile as a 2D Knudsen field and show that
a unified treatment, including continuum and non-continuum mechanics, is required to
comprehend the air layer dynamics coherently. The airflow dynamics in the central dimple
region falls within the purview of continuum Stokes regime. In contrast, the peripheral
air disc falls within the non-continuum slip flow and transition regime characterized by
high Knudsen number values. However, the initial average air disc expansion dynamics
could be understood in terms of Stokes approximation. In non-continuum regimes of
the peripheral air disc, we discover intriguing interface perturbations that leads to
asymmetric wetting of the substrate. The first point of contact between the drop and
substrate occurs at the edge of the peripheral disc region. The perturbative structures in
the disc region exist for a short time due to the asymptotic effects of capillary and van
der Waals interaction at the sub-micron length scales.
\end{abstract}

\begin{keywords}
Authors should not enter keywords on the manuscript, as these must be chosen by the author during the online submission process and will then be added during the typesetting process (see http://journals.cambridge.org/data/\linebreak[3]relatedlink/jfm-\linebreak[3]keywords.pdf for the full list)
\end{keywords}

\section{Introduction}
Drop impact phenomena is quite ubiquitous and spectacular. Appreciating the elegant structures that appear during drop impact dynamics on solids and liquids does not require scientific training. This phenomenon thus appears in a variety of advertisements/artistic contexts in the media industries. The physics underlying the aesthetic beauty of drop impact phenomena is equally beautiful and attracts the scientific community to study it in various contexts. The original study of drop impact physics began with the seminal work of Worthington \citep{worthington1877xxviii,worthington1877second,worthington1883impact} in the late nineteenth century. Using ingenious mechanical arrangements to synchronize the fall of the drop with that of an electric discharge, Worthington was able to study drop impact phenomena on solids and liquids. Initial observations made by Worthington were manual and hand drawn. Later on, Worthington was able to record the impact phenomena on photographic plates \citep{worthington1897v,worthington1900iv}. 

With the advent of high-speed imaging, experimental drop impact physics was revolutionized as scientists and engineers could probe smaller and finer spatio-temporal scales \citep{thoroddsen2008high,yarin2017collision}. Experimental drop impact research was majorly driven by various technologies and industrial processes like inkjet printing \citep{wijshoff2018drop,lohse2022fundamental}, food processing industries \citep{andrade2013drop}, spray technologies \citep{breitenbach2017heat,breitenbach2018drop} to mention a few. The landscape of outcomes from any drop impact scenario are enormous.
The variety of impact outcomes depends on the impact geometry, substrate type (solid vs. liquid), ambient pressure, and surface temperature \citep{yarin2006drop,xu2005drop,roy2019dynamics,roy2023mechanics}. 
A general drop impact phenomenon can be characterized experimentally by a set of non-dimensional numbers like Weber number, Reynolds number, Ohnesorge number, Bond number, Capillary number, and Mach number, to name a few. Each of these non-dimensional numbers represents the balance of dominant scales related to a particular phenomenon observed during impact events. Using appropriate non-dimensional quantities various kinds of impact events like solid impact, drop impact, jet impact 
\citep{roy2022thermofluidics}
to name a few could be understood in a unified framework \citep{yarin2017collision}. Such unified description can explain a wide range of impact phenomena ranging from meteor impacts to granular matter impacts to jet impacts on solids 
and thin films like bio-mechanical systems \citep{dowling2013scaling,roy2021fluid,gilet2015fluid,wang2018non,bourouiba2021fluid}. Some phenomena related to drop impacts are spreading, splashing, bouncing, corona formation, and various kinds of Worthington jets. These phenomena are very well studied experimentally \cite{yarin2017collision,thoroddsen2012micro} especially spreading and splashing due to their wide application in industries. Recently, scientists are also studying  impact phenomena from infectious disease and pathogen transmission perspective
\citep{shetty2020quantitative,roy2021fluid}
\citep{Hariharan2022.05.28.493826}. 
Drop impact generally is multi-scale, ranging from nano-scale \citep{li2015probing,de2015wettability} to millimeter scales. Despite immense research in drop impact studies for the last one and a half-century, several aspects of drop impact physics still remain elusive and need further investigation.

One such area is the air layer dynamics beneath an impacting drop. When a drop approaches an impacting interface, 
 the bottom part (leading edge) of the drop experiences a rise in pressure due to lubrication pressure buildup. The lubrication pressure, in general, is inversely proportional to the distance between the leading edge and the impacting surface. The increased pressure at the leading edge of the drop causes a 
 deformation of leading edge of the drop, entrapping air sandwiched between the the drop and the impinging interface/substrate.
 As the air between the drop and substrate thins out various interactions like capillary and van der Waals becomes dominant. The stability of the air layer between the drop and the impacting surface depends on the interaction between capillary and van der Waals interaction. 
For high speed impact, inertial and compressible effects will also be important \citep{lesser1983impact} along with the capillary and van der Waals interaction effecting the air layer dynamics. 
 Recent experimental works have shown that drops in the size range of millimeters are supported by micro and nano air films \citep{kolinski2012skating,langley2017impact,langley2019gliding}. The destabilization of the air layer is inherently related to bubble formation in various drop impact systems. Bubbles in industrial and various engineering impact systems like printing, painting, and cooling are often detrimental to the optimum working efficiency of the system. However, bubbles could also be helpful, as is seen in various natural systems like raindrops impacting liquid bodies. In these natural systems, bubbles help in gaseous exchange between the atmosphere and the liquid bodies supporting aquatic/marine life forms \citep{woolf2007modelling}.

On approaching the impacting surface, the drop experiences a pressure rise at the leading edge of the drop due to air drainage (outward radial flow) in between the drop and the substrate. 
Based on the relative air layer thickness, the entrapped air layer can be subdivided into two distinct disjoint regions; the central dimple and the peripheral air disc region \citep{roy2022droplet}.
The dimple forms beneath the impacting drop when the lubrication pressure exceeds the capillary pressure. This is in contrast to solid particle impact at low to moderate velocities, where deformation of the impacting solid object is negligible since the particle behaves as a rigid body. 
The actual contact between the impacting objects occur, when the dimensions of the entrapped draining air film reaches and becomes smaller than the mean free path of non continuum molecular regimes.
Non-continuum analysis in lubrication drainage flows was incorporated by Hocking \citep{hocking1973effect}. Hocking modified the standard lubrication analysis by using the Maxwell slip boundary condition instead of the no-slip boundary condition. Using Hockings analysis as a base, rigid particle impacts like the squeeze flow problem (between rigid spheres or a sphere and a flat plate) typically found in gas-solid suspension, were studied in extensive detail \citep{davis1984rate}. The monodisperse gas-solid suspension was studied mathematically using classical and non-continuum lubrication flows by many scientists \citep{koch1990kinetic,sundararajakumar1996non,how2021non}. In this work we use some of the asymptotic methods 
and concepts
developed in the context of 
entrapped air layer dynamics during
solid particle impact.
The central dimple that forms in drop impact systems has been explored by scientists in the last decade experimentally and computationally \citep{hicks2010air,hicks2011air,de2012dynamics,de2015wettability,de2019predicting,driscoll2011ultrafast}.
However, surrounding the central air dimple exists a region that we refer to as the peripheral air disc \citep{roy2022droplet}.
The air layer dynamics in the peripheral region remain relatively unexplored, and scientific literature is sparse. 
Some of the important works in the context of drop impact exploring the central dimple experimentally and computationally are van der Veen et al. \citep{van2012direct}, Bouwhuis et al. \citep{bouwhuis2012maximal}, Mandre et al \citep{mandre2009precursors,mandre2012mechanism}. 
The work by van der Veen et al. \citep{van2012direct}, is an experimental study that measures the air layer layer profile under impacting drop using high speed color interferometry. The major focus was to measure the time evolution of the air layer profile before and during the contact process with the substrate. Further, from the air layer profile, the velocity of the air exiting the gap between the drop and the substrate during the wetting and bubble formation process was calculated. 
Bouwhuis et al. \citep{bouwhuis2012maximal}, explored the competing mechanisms that contols and minimizes the entrained air bubble size during drop impacts on solid surfaces. For large impact size and velocity, the bubble is flattened due to the inertia of the droplet. For smaller size and velocity, capillary effects minimizes the droplet size. Bouwhuis et al. \citep{bouwhuis2012maximal} therefore showed that the entrained bubble volume can have a maximum size under the effect of inertial and capillary action. Mandre et al. \cite{mandre2009precursors} studied the splashing dynamics during drop impact on a solid surface and shows that the initial spreading of the droplet occurs on top of a thin air layer that prevents the initial wetting of the droplet and the substrate. During the initial spreading phase, the drop-air bottom interface develops regions of very high curvature which further generates capillary waves. Another study by Mandre et al. \citep{mandre2012mechanism} provides an ab-initio description of sheet formation during drop impacts on dry solid surfaces. Further, the major part of the study by Mandre et al. was about the liquid phase and the prediction of sheet formation as a function of experimental parameters like drop size and velocity. The sheet formation dynamics also provides a semi-quantitative method to study various processes related to the sheet formation dynamics and understand various parameter dependence like viscosity, gas pressure, surface roughness, and surface tension. Josserand et al. \cite{josserand2016drop} wrote an excellent review article that provides a state of the art research for drop impacts on solid surfaces. The article also discusses some of the work related to the measurement of the central air dimple profile. It is to be noted that the above mentioned works does not discuss anything related to non-isothermal impacts, Knudsen field, structures in the peripheral disc, which are some of the major contribution of the current work. In this work, we study the entire air layer profile beneath the impacting droplet on heated substrate experimentally and theoretically under a unified framework using high speed reflection interferometry imaging, asymptotic methods and scaling analysis. Further, the present study also highlights the need of non-continuum hydrodynamics to properly understand the complete picture. We show experimentally how the air layer profile consists of regions that encompasses the entire regime from continuum to non-continuum hydrodynamics beneath the impacting droplet impinging on heated substrate.
Recently Chubynsky et al. \citep{chubynsky2020bouncing} shed some light on the effect of van der Waals force related to gas kinetic effects (GKE) in drop impacts for low Weber number computationally.

In our previous work \citep{roy2022droplet}, we have explored the peripheral air disc in the context of drop impact on immiscible liquid pools. We discovered that new structure could exist in the peripheral disc due to balance of capillary and van der Walls (vdW) interaction.
This work focuses on the air layer dynamics during drop impact on heated substrates at low impact energy/Weber number below the Leidenfrost temperature. We study the dynamic evolution of the central air dimple and the peripheral air disc for different substrate temperatures ranging from $300$K to $473$K. The impact Weber number ($We={\rho}_l{V_0}^2R_0/{\sigma}_{aw}$) based on drop radius $R_0$, liquid density ${\rho}_l$, impact velocity $V_0$, and air-water surface tension ${\sigma}_{aw}$ was kept low ($We{\sim}1$) so that the effect of inertia and compressibility could be neglected. At low Weber number regime, lubrication approximation holds good. We 
have detected 
non-isotropic structures in the peripheral air disc using high-speed reflection interferometric imaging. The structures in the peripheral air disc region cause asymmetric wetting between the drop and the substrate.  
The film mode and kink mode are the two fundamental contact modes designated by some recent works \citep{de2015air-1,de2015air-2,li2015probing,langley2019gliding}. The film mode, or the first kink mode, occurs just outside the dimple, whereas the kink mode, or the second kink mode, occurs at the maximum extension of the drop \citep{chubynsky2020bouncing}. 
The film mode occurs for $We>>1$ whereas the kink mode occurs for $We{\sim}1$. Based on our previous \citep{roy2022droplet} and current work, we propose that the underlying cause of both the film and kink mode are the non-isotropic structures in the peripheral air disc region.

As the drop approaches the heated impacting surface 
(refer to Fig. 1 for the experimental impact configuration), i.e., the gap between the bottom most point of the drop and the surface decreases (Fig. 2(a)), the lubrication pressure
increases (since the lubrication pressure is inversely proportional to the gap between the drop and the surface). The lubrication pressure when exceeds the local capillary pressure at the bottom most point of the droplet results in the formation of a dimple. The air-water interface at the bottom of the drop therefore is essentially a curved surface entrapping a thin layer of air having thickness ranging from  $h{\sim}5-10{\mu}$m at the center to $h{\sim}70-100$nm at the edge. The air layer profile beneath the drop hence, in general can be thought of as consisting of two disjoint regions based on the relative air layer thickness $h$, the central air dimple, $h{\sim}5-10{\mu}$m(refer to '*' in Fig. 2 (b) depicting the central air dimple) and the surrounding region around the central dimple that the authors refer to as the peripheral air disc, $h{\sim}70-100$nm (refer to '\#' in Fig. 2 (b) and 2(c) for a schematic of the peripheral air disc and a 3D non-dimensional height profile respectively).
The central air dimple has been generally studied for iso-thermal drop impact conditions both experimentally and theoretically. It is important to note that non-isothermal drop impact has also been studied extensively by the drop impact community. The major focus though were generally the experimental characterization of impact processes in various regimes like spreading, recceding, leidenfrost, bouncing, and atomization for various non dimensional parameters like Weber number, Capillary number, Reynolds number, Ohnesorge number, Stokes number, viscosity ratio and density ratio between the drop and the surrounding gas. However, there are relatively few studies that explores the central air dimple under non iso-thermal drop impact conditions using high speed interferometry \citep{qi2020air}. It is important to note that the central air dimple thickness is at least one order of magnitude larger than the peripheral air disc thickness. The peripheral air disc region thus is very sparsely studied and still remains elusive. The peripheral disc region is important as the first point of contact generally occurs in this region and is essential for an in-depth quantitative understanding of the bubble formation process. The first point of contact in drop impact scenarios were previously reported to occur majorly in two modes, the first kink mode or film mode for large impact Weber numbers and second kink mode or kink mode for Weber number of the order of unity. In this work we generalize based on the current and our previous impact experiments on liquid and solids that the first and second kink mode are a special case of various structures that are present in the peripheral air disc region as a result of the asymptotic balance between capillary and van der Waals forces. These structure acts as nucleation point of air film destabilization and form rupture points that subsequently leads to air film dewetting and bubble formations. Fig. 2(d) shows a typical Knudsen field beneath an impacting droplet. Two distinct regions are clearly visible, the central dimple having relatively small value of Knudsen number ($Kn<0.01$) can be understood using continuum hydrodynamics. On the contrary the peripheral air disc region is not smooth and has structures of various length scales. The Knudsen number in the disc region is relatively large ($0.01<Kn<0.1$) and hence falls within the purview of slip flow and non-continuum hydrodynamics regime where the first point of contact occurs.
\color{black}

The present study explores the air layer dynamics at low-impact Weber number regime ($We{\sim}1$) on heated glass substrate below Leidenfrost temperature using high speed reflection interferometry imaging and theoretical analysis. The main theme of this work is to understand the mechanics of the central air dimple and its immediate neighbourhood (the peripheral air disc). The detailed characterization for various experimental parameter ranges is outside the scope of the present work.
\begin{figure*}
    \centering
    \includegraphics[scale=1]{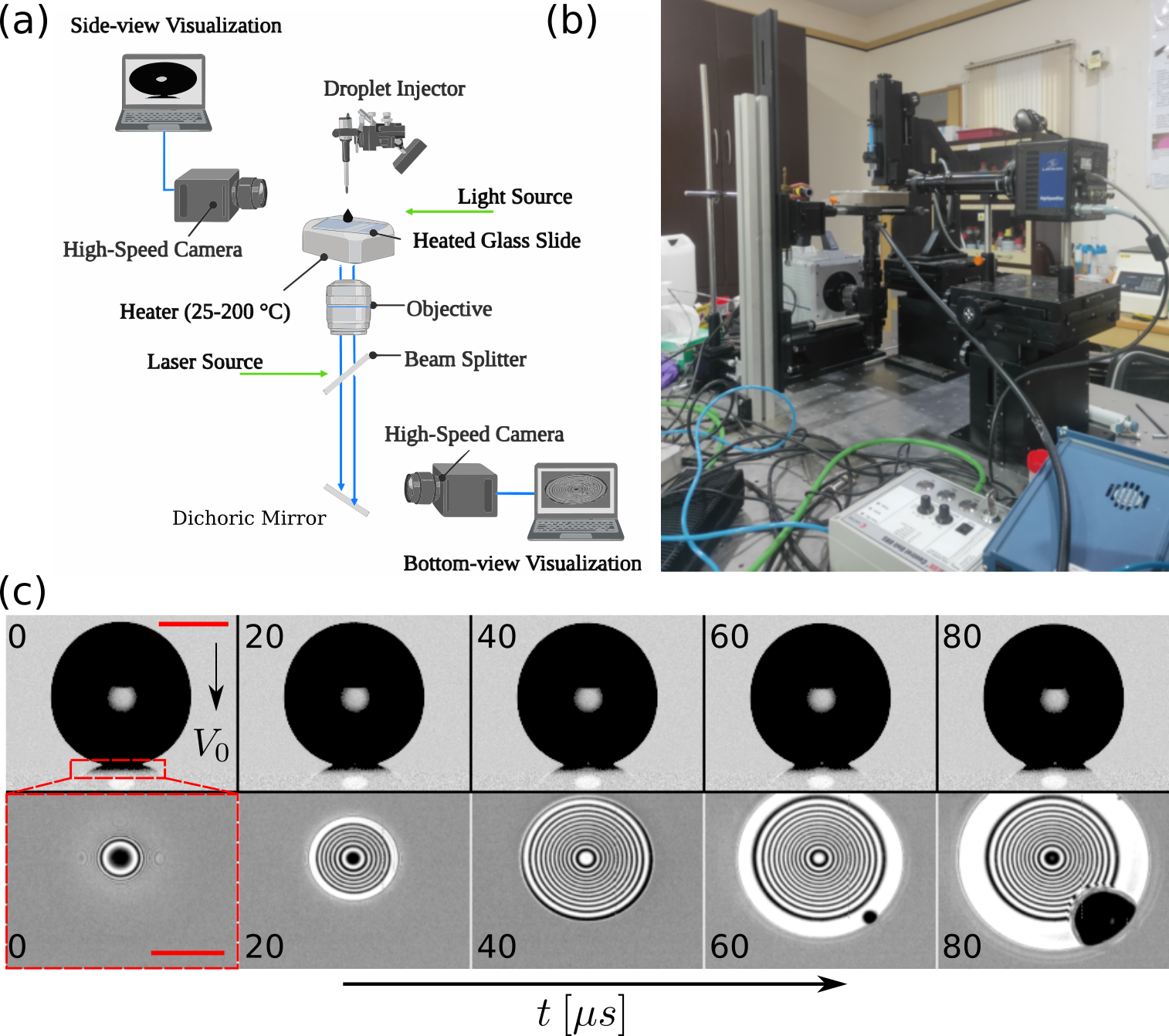}
    \caption{(a) Schematic of the experimental set up. (b) Actual experimental setup. (c) A typical time series depicting the side and interferometric view of an impacting drop on a substrate with temperature $T_s=300$K respectively. Timestamps are in microseconds and the scale bar for the side view and interferometric view represents $1.3$mm and $0.296$mm respectively.
    }
    \label{Figure1}
\end{figure*}

 \section{Materials and Methods}
 We study liquid water drop impact on heated solid glass substrate below the Leidenfrost temperature.
 The drop on approaching the impacting surface causes lubrication pressure rise in the thin 
entrapped air layer between the drop and the substrate. The lubrication pressure on exceeding the capillary pressure across the drop interface results in the formation of the central air dimple and the peripheral air disc. We study evolution of the central air dimple and the peripheral air disc using high speed imaging and theoretical analysis.
 \subsection{Experimental Set up}
Using high-speed reflection interferometric technique, we study air layer dynamics beneath the impacting drop on a glass substrate at various surface temperatures. The schematic and assembly of the actual experimental setup is shown in Fig. 1(a) and 1(b), respectively. De-ionized water drop of radius 
$R_0=1.1\pm0.1{\:}$mm
generated from a syringe pump (New Era Pump Systems, NE-1010) impinges a glass substrate of dimensions 
$75\times25\times1$mm$^3$
(procured from Blue Star) with an impact velocity of 
$V_0=0.25$m/s ($We{\sim}1$). The corresponding impact energy of the drop is $0.3{\mu}$J.
The drop was allowed to fall freely under gravity from a designated height to achieve the desired impact velocity.
We used cleaned glass slides for the experiments. The glass slides were sonicated in a bath of isopropanol for 10 minutes. The glass slide was then placed on the annular-shaped heater (copper surface) connected to a PID controller for maintaining the precise temperature conditions on the heated surface. The surface temperature of the glass plate at the impact region (placed on the annular heater) was measured using a K-type thermocouple within the error range of 
$\pm5^{\circ}$C=$\pm5$K. 
We capture the drop impact phenomena using two simultaneous high speed imaging methods (shadowgraphy and interferometry) at identical frame rates to understand and characterize the impact and air-layer dynamics for substrate temperatures $300$K, $353$K, $423$K and $473$K. Shadowgraphy is performed from the side view, and reflection interference imaging from the bottom view as shown in Fig. 1(a) and Fig. 1(b). For the shadowgraphy we use a LED (Light Emitting Diode) light source for backlight and a high speed camera connected with a zoom lens (Navitar). The shadowgraphy 
and the interferometry imaging were both
performed at 50000 FPS (frames per second). Sample shadowgraphy and interference image sequence are shown in the top and bottom image sequence panel of Fig. 1(c) respectively for substrate temperature 
$T_s=300$K.
The drop impacts the glass with a velocity 
$V_0=0.25$m/s.
The dotted rectangle in the side view shows the field of view for the bottom view interferograms.
The timestamps are in microseconds for both the shadowgraphy and interference image sequence. The scale bar shown in red for the side and interferometric views represent $1.3$mm and $0.296$mm respectively.  
The reflection interference imaging setup was assembled on the bottom side of the glass plate to record the dynamic interference patterns during drop impact. The interference system consists of a pulse-diode laser source (Cavitar Cavilux smart UHS, 400 W power, ${\lambda}_s=640$nm) for illumination purpose; a beam-splitter; a 4x microscope objective; a dichoric mirror; and a high speed camera as shown in Fig. 1(a). The reflected light beam from the beam-splitter passes through the microscope objective onto the glass surface. The reflected light from the top glass surface and the bottom surface of the air-water interface of the impacting drop superimposes to produce an interference pattern. The interference pattern was focused using a zoom lens (Navitar) with the assistance of in-line dichoric mirror, as shown in the 
schematic of Fig. 1(a) and the actual experimental setup Fig. 1(b).
The dynamic fringe patterns were recorded using a high-speed camera (Photron SA5) with similar frame rates to the side-view imaging (i.e., at 50000 FPS). A pixel resolution of 
$1.5{\mu}$m
per pixel was used to resolve the constructive and destructive interference patterns. 
\subsection{Interferometry post-processing} 
A monochromatic beam from the diode laser source passes 
from the bottom to the top
through the microscope objective and gets transmitted through the glass substrate (Fig. 1(a)). Some part of the transmitted beam suffers reflection at the air-glass interface (top surface of the glass substrate) and travels back to the CCD of the high-speed camera with an intensity 
$I_1(x,y)$ where $(x,y)$ represents the image coordinate.
The other part of the transmitted beam coming out of the glass substrate travels through the air layer beneath the impacting drop and undergoes reflection at the air-water 
drop
interface (bottom surface of the impacting drop) (refer to Fig. 2(a) and 2(b)). The reflected beam from the 
air-water drop
interface travels back to the CCD of the high-speed camera with an intensity
of $I_2(x,y)$.
Beams with intensity 
$I_1(x,y)$ and 
$I_2(x,y)$
superimpose to produce interference pattern recorded by the high-speed camera. The interference occurs, because the beams 
$I_1(x,y)$
and 
$I_2(x,y)$
are coherent. Further, the beam with intensity 
$I_2(x,y)$
is phase shifted to the beam with intensity 
$I_1(x,y)$
due to the additional path traversed through the air layer. The path difference between beams 
$I_1(x,y)$
and
$I_2(x,y)$
causes a phase shift producing a resultant interference pattern of intensity 
$I(x,y)$
given by \citep{daniel2017oleoplaning,limozin2009quantitative,sugiyama2006single}
\begin{equation}
I(x,y) = I_1(x,y)+I_2(x,y)+2\sqrt{I_1(x,y)I_2(x,y)}cos({\phi}(x,y))
\end{equation}
where 
${\phi}(x,y)$
is the phase difference between beams having intensity 
$I_1(x,y)$
and 
$I_2$(x,y).
$I(x,y)$ is the resultant intensity of the interference pattern recorded by the camera at an image coordinate $(x,y)$. Sample dynamic side views image sequence and interferograms are shown in Fig. 1(c) during drop impact on 
$300$K
glass substrate at $We{\sim}1$. The timestamps are in microseconds.
The main feature of equation (2.1) is to represent the resultant intensity distribution on the image coordinate plane due to interference of beams having intensity 
$I_1(x,y)$, $I_2(x,y)$
and the cosine-modulated phase field given by $cos{\phi}(x,y)$. 
If the phase gradient is high, high-density fringes are observed. Similarly, a smaller phase gradient corresponds to low fringe density. The dark fringes correspond to destructive interference, whereas the 
bright
fringes correspond to constructive interference. 
From equation (2.1), we can observe that a general resultant intensity field could be represented mathematically as
\begin{equation}
    I(x,y)=B(x,y)+A(x,y)cos({\phi}(x,y))
\end{equation}
where $B(x,y)=I_1(x,y)+I_2(x,y)$ and $A(x,y)=I_1(x,y)I_2(x,y)$. Physically $A(x,y)$ and $B(x,y)$ represents the fringe amplitude and background intensity fields respectively. On proper background subtraction and fringe amplitude normalization, the resultant intensity field $F(x,y)$ could be represented as
\begin{equation}
    F(x,y)=\frac{I(x,y)-B(x,y)}{A(x,y)}=cos({\phi}(x,y))
\end{equation}
Therefore from an input experimental image $I(x,y)$, a background subtracted normalized image $F(x,y)$ could be computed using equation (2.3). The unknown phase field ${\phi}(x,y)$ could be computed from the inverse cosine function, i.e.,
\begin{equation}
    {\phi}_0(x,y)=cos^{-1}(F(x,y))
\end{equation}
The inverse cosine function being multi-valued, the true phase ${\phi}(x,y)$ could be determined using a proper signature of ${\phi}_0(x,y)$ if represented in a wrapped form ($\phi(x,y)=\pm{\phi}_0(x,y)$).
Given that $-1{\leq}F(x,y){\leq}1$, ${\forall}$ $(x,y)\in[0,x_{max}]{\times}[0,y_{max}]$, i.e. for all $(x,y)$ coordinates in the image plane, $\phi_0(x,y)\in[0,\pi]$ for single valuedness of the inverse cosine function. Here the pair $(x_{max},y_{max})$ represents the region of interest boundary coordinates diagonally opposite to $(0,0)$ which is chosen as the origin in the image plane. For the entire image captured from the CCD sensor, the pair $(x_{max},y_{max})$ could be mapped to the image resolution in appropriate length units. In general, frequency guided sequential demodulation (FSD) methods \citep{kemao2007sequential,wang2009frequency}, the actual phase field is calculated using linear and quadratic approximation methods, where the gradient phase fields (also known as local frequencies) are calculated by optimization an appropriate cost function using least square methods. A
typical cost function that can be used for the optimization process and computing the gradient fields $({\omega}_x(x,y),{\omega}_y(x,y))$ is given by
\begin{equation}
    C(x,y)=\sum_{(\delta_1,\delta_2){{\in}N_{xy}}}[F({\delta}_1,{\delta}_2)-cos({\phi}(x,y;\delta_1,\delta_2))]^2
\end{equation}
where $N_{xy}$ denotes a neighborhood around the point $(x,y)$.
Assuming linear approximation to compute ${\phi}(x,y;{\delta}_1,{\delta}_2)$ in the neighborhood $N_{xy}$ from ${\phi}_0(x,y)$, we have using the linear Taylor's series approximation in two dimensions
\begin{equation}
    {\phi}(x,y;{\delta}_1,{\delta}_2)={\phi}_0(x,y)+{\omega}_x(x,y)({\delta}_1-x)+{\omega}_y(x,y)({\delta}_2-y)
\end{equation}
The proper sign of the phase field is determined by allowing the gradients to be continuous. The optimization is generally done using exhaustive search algorithms and more recently using Levenberg-Marquardt optimization methods to improve the computational time \citep{wang2009frequency}.
In this work, the dynamic fringe patterns during drop impingement have been processed using the fast-frequency guided sequential demodulation (FFSD) method \citep{kai2010fast,wang2009frequency,roy2022droplet}. Fast-frequency guided algorithms are suitable for transient phenomena \citep{roy2022droplet} and consequent comprehensive extraction of the two-dimensional phase field of the complex fringe patterns. The FFSD algorithm has been used to calculate the signature of the unknown phase. Using equation (2.3) we can construct a scalar field that will be used for signature determination as
\begin{equation}
    \bm{{\nabla}}F(x,y){\cdot} \bm{{\nabla}}{\phi}(x,y)=-sin({\phi}(x,y))|\bm{{\nabla}}{\phi}(x,y)|^2
\end{equation}
$\bm{{\nabla}}F(x,y)$ can be computed from the field $F(x,y)$ and $\bm{{\nabla}}{\phi}(x,y)$ can be computed using ${\phi}_0(x,y)$ from equation (2.4) in the neighborhood $N_{xy}$ around the point (x,y). From equation (2.7) we can see that the sign(${\phi}(x,y)$)=-sign($\bm{{\nabla}}F(x,y){\cdot}\bm{{\nabla}}{\phi}(x,y))$ since $|\bm{{\nabla}}{\phi}(x,y)|^2{\geq}0$ and sign($sin({\phi}(x,y))$)=sign(${\phi}(x,y)$). Therefore, to determine the signature of the $\phi(x,y)$, we have to determine the signature of $\bm{{\nabla}}{\phi}(x,y)$.
The gradient of phase distribution for the obtained fringe patterns 
could be represented in terms of complex numbers for efficient computing as
\begin{equation}
    \bm{\nabla}{\phi}_0(x,y)={\omega}(x,y)e^{i{\theta}(x,y)}
\end{equation}
Here, ${\omega}(x,y)$ signifies the total amplitude 
of the phase gradient field vector and ${\theta}(x,y)$ represents the direction of the vector field at any point $(x,y)$. The complex field from equation (2.8) is smoothed by averaging the square of the gradient phase field values over some neighborhood region $N_{xy}$ to obtain ${\omega}_s(x,y)$ and ${\theta}_s(x,y)$. The suffix 's' denotes the smoothed field. Using smoothed version of the phase gradient field, the phase gradient field could be computed without any sign ambiguity by requiring spatial continuity on the gradient of the phase field, as $\bm{\nabla}{\phi}(x,y)=\bm{\nabla}{\phi}_{0s}(x,y)$ if $\bm{\nabla}{\phi}(x,y){\cdot}\bm{\nabla}{\phi}_{0s}(x_*,y_*){\geq}0$ where $(x_*,y_*)$ denotes the adjacent pixels of $(x,y)$. Similarly, $\bm{\nabla}{\phi}(x,y)=-\bm{\nabla}{\phi}_{0s}(x,y)$ if $\bm{\nabla}{\phi}(x,y){\cdot}\bm{\nabla}{\phi}_{0s}(x_*,y_*){<}0$. Using the proper sign of the gradient of the phase field in equation (2.7), the phase field signature can be determined uniquely without any ambiguity. However, the phase field can be ambiguous to a factor of $2k{\pi}$ in general (the period of cosine modulated function), where $k{\in}Z_{I}$, $Z_{I}$ represents set of all integers. To circumnavigate the phase ambiguity issue,
the obtained phase are constructed to be continuous and unwrapped. 
Using the unwrapped phase field ${\phi}(x,y)$, the two dimensional thickness profile $h(x,y)$ could be computed. The phase ambiguity can be handled using various interferometry techniques like color interferometry, dual/multi wavelength interferometry \citep{josserand2016drop}. Further, the thickness profile can also be measured without a $2{\pi}$ phase ambiguity using monochromatic interferometry. The fringes obtained from monochromatic and coherent source of illumination is known as Haidinger fringes \cite{hecht2012optics} in general. Some of the commonly used techniques to determine the thickness without a phase ambiguity for the monochromatic illumination are the peak and valley methods \cite{gillen2005use,shukla2006non}, reference thickness methods \cite{bouwhuis2012maximal}, Haidinger quadrature \cite{kim2017interferometric,park2019review} to name a few. We use the reference thickness methods in our measurements to determine the absolute thickness profile, since the absolute thickness are known at certain positions of the interference images.
One such reference point is the hole nucleation spots (air layer rupture locations) where the absolute thickness monotonically approaches zero in time (refer to '*' in Fig. 1 of the supplementary document) and hence any arbitrary integer multiple of wavelength will not be able to match the actual thickness profile evolution close to the rupture (nucleation) locations. Therefore temporal causality for the evolution of the thickness profiles at the hole nucleation point can lead to the determination of absolute thickness from a monochromatic illumination.
The evolution of air layer thickness in general is a causal process in time and hence allow the computation of actual air layer thickness based on the rupture locations, where the thickness profile goes to zero provided the phase field is continuous. Further, based on the observations in the current experiment, that most rupture points are very close to the outer expanding rim, the outer rim location are used to set up a zero thickness reference for very early time data when rupture locations have not formed (refer to '*' in Fig. 1 of the supplementary document). It is important to observe that the average air layer thickness decreases with radial coordinate upto the outer expanding rim of the air disc. The average thickness at the outer expanding rim therefore will be smallest in the the entire air layer region well beyond the measurement range and hence can be considered as reference points of zero thickness.\\\\
The height of the air-layer profiles can 
then
be estimated for the preferred wavelength of the light source ${\lambda}_s$ and refractive index $n$ of the fluid medium as \citep{kitagawa2013thin}:
\begin{equation}
    h(x,y)=\frac{{\lambda}_s}{2{\pi}n}{\phi}(x,y)
\end{equation}

The images acquired with the shadowgraphy technique (i.e., side-view imaging) were processed using a combination of open-source tools like ImageJ \citep{schneider2012nih} and python \citep{10.5555/1593511} for one to one comparison for variation of drop shape during impact and simultaneous change in interference pattern for varying surface temperatures. Sufficient number of experimental runs were conducted to ensure statistically significant datasets.
\section{Results and Discussions}

\color{black}
\subsection{Global overview}
Fig. 2(a) depicts the schematic and the geometry of the impact condition under study. The air layer dynamics beneath an impacting drop on a heated substrate at various substrate temperatures
($T_s=300$K, $353$K, $423$K and $473$K)
has been studied using high-speed reflection interferometry imaging at low impact kinetic energy ($We{\sim}1$). 
The air layer beneath the drop is subdivided into two distinct disjoint regions that we refer to as the central air dimple, surrounded by the peripheral air disc (Fig. 2(b)). Fig. 2(b) depicts the air layer profile schematically. The * denotes the central air dimple, and \# represents the peripheral air disc. The thickness of the central dimple, in general, is larger than the peripheral air disc region by one order of magnitude (i.e., $h_{*}/h_{\#}{\sim}\mathcal{O}(10)$). We observe using high-speed interferometry imaging that the geometrical features features of the dimple (i.e., central dimple maximum thickness and central dimple diameter) are weakly dependent on the substrate temperature $T_s$ and depends on the impact energy/Weber number ($We$) in general. The average maximum height of the central air dimple is
${\sim}\mathcal{O}$($6{\mu}$m).
We also observe that the air layer rupture time scale and rupture radius increases with increase in the substrate temperature. The rupture time scale lies within the range of
${\sim}\mathcal{O}$($100{\mu}$s) to ${\sim}\mathcal{O}$($400-500{\mu}$s)
for a substrate temperature variation between 
$300$K to $473$K
respectively. Similarly, the rupture radius variation from 
$300$K to $473$K
substrate temperature lies within the range 
${\sim}\mathcal{O}$($200{\mu}$m) to ${\sim}\mathcal{O}$($600{\mu}$m)
respectively. 
Further, we show that a Gaussian profile can approximate the shape of the central air dimple during its 
temporal
evolution. A standard continuum Stokes flow regime governs the central dimple evolution. In contrast, the dynamics in the peripheral disc region falls in the non-continuum regime characterized by a relatively high value of Knudsen number (Fig. 2(c) and 2(d)). Fig. 2(c) shows a 3d wireframe plot of the non-dimensional height profile of air layer at a surface temperature of 
$T_s=423$K
normalized with respect to the mean free path of air. The non-dimensional height profile is a one-to-one map of the inverse Knudsen field map. The dotted 
red
rectangle in Fig. 2(b) and Fig. 2(c) depicts the interfacial perturbations in the peripheral air disc. Fig. 2(d) depicts the Knudsen contour field corresponding to Fig. 2(c). We observe intriguing 3d structures along the peripheral air disc region that acts as a seed for asymmetric wetting by the drop on the substrate. Such structural length scales exist at short time scales due to balance of capillary and molecular interactions like van der Waals. 
The
structures in the peripheral air disc region lead to asymmetric wetting between the drop and substrate. The structures in the peripheral disc are the underlying cause of the kink and film mode of first contact between the drop and substrate. For the impact Weber number in this study ($We{\sim}1$), we observe kink mode of contact in contrast to the film mode of contact observed in our previous study on liquid pools ($We{\sim}10$) \citep{roy2022droplet}. This is in accordance with the numerical findings of Chubynsky et al. \citep{chubynsky2020bouncing}.

 The remaining article is arranged as follows. From section 3.2 onwards, we analyze the air layer dynamics using experimental and theoretical methods. We subdivide the analysis into various sub-sections. Section 3.2 introduces the coordinate system and the initial condition before dimple formation. Section 3.3 focuses on the asymptotic analysis for the velocity and pressure field in the dimple region. Section 3.4 provides an estimate of the dynamic dimple radius. Section 3.5 deals with the radial expansion of the peripheral air disc. We discuss and explore the air layer dynamics and perturbative structures in the peripheral air disc in section 3.6.

We begin the analysis by considering the geometry and initial condition for understanding air layer dynamics underneath the impacting drop. 

\subsection{Coordinate system and Air profile just before dimple formation}
\begin{figure*}
    \centering
    \includegraphics[scale=1]{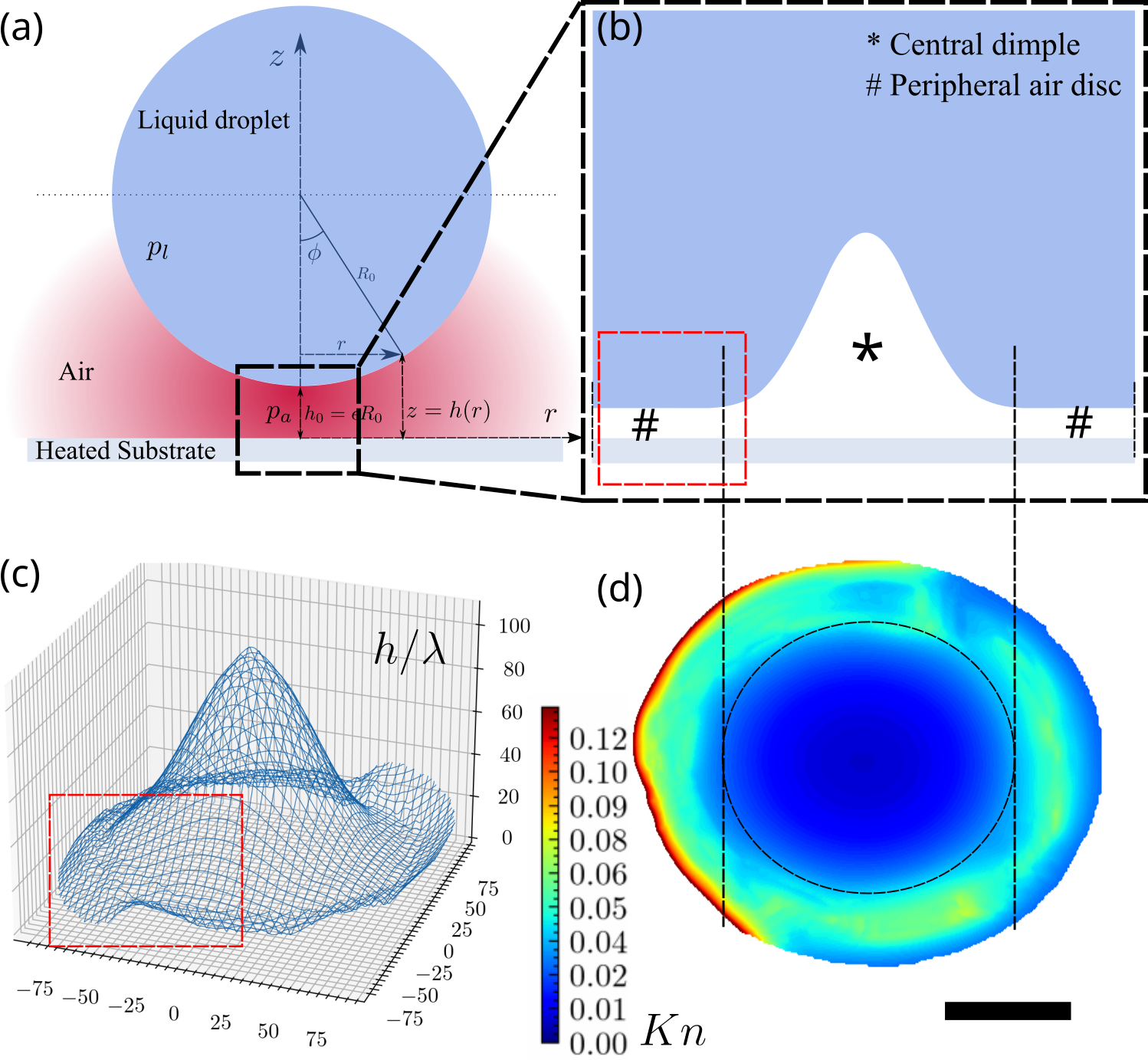}
   \caption{(a) Schematic representation depicting the initial impact configuration, geometry and coordinate system just before dimple formation occurs. (b) Close up schematic of the bottom most point of the drop depicting the central air dimple (*) and the peripheral air disc (\#). (c) A typical 3D non-dimensional air layer thickness profile ${h/{\lambda}}$ depicted as a wireframe plot for substrate temperature of $T_s=423$K depicting the interface perturbations in the peripheral disc region. $h$ represents the dimensional air layer thickness and ${\lambda}$ is mean free path of air. (d) A typical Knudsen field contour map of the air layer beneath the drop for substrate temperature of $T_s=423$K corresponding to panel (c). The black scale bar represents $195{\:}{\mu}$m and the color bar represents the Knudsen number value $Kn$.}
    \label{Fig2}
\end{figure*}

Fig. 2(a) depicts a schematic representation of the impact geometry just before the central air dimple begins to form. We use a cylindrical coordinate system to pose the mathematical problem.

From the geometry of the drop, using elementary trigonometric correlations (Fig. 2(a)) we have
\begin{equation*}
    sin{\phi} = \frac{r}{R_0};{\:}cos{\phi}=\frac{\sqrt{R_0^2-r^2}}{R_0}
\end{equation*} 
where ${\phi}$ is an angle parametrizing the bottom surface of the drop, $r$ is the radial coordinate from the axis of the cylindrical coordinate system (vertical $z$ axis), and $R_0$ is the impacting drop radius. The build of lubrication pressure in the air layer just before the dimple formation is shown schematically
as a red halo in Fig. 2(a).
The pressure in the liquid drop and the air layer is $p_l$ and $p_a$ respectively. The closest approach distance between the drop and the substrate is denoted by $h_0={\epsilon}R_0$, where ${\epsilon}$ is a small asymptotic parameter that we will be used later in computing the pressure and velocity field respectively.   
The bottom part of the impacting drop just before the dimple formation can be approximated as part of a sphere (Fig. 2(a)). The initial air layer thickness profile therefore can be written as
\begin{equation}
    z(r) = h_0 + R_0\left(1-\sqrt{1-\left(\frac{r}{R_0}\right)^2}\right)
\end{equation}
where $z(r)$ is the air thickness at any radius $r$ between the drop and the substrate. $h_0$ is the minimum air layer thickness between the drop and the substrate before dimple formation begins to occur at $r=0$. Being part of a sphere, equation (3.1) is symmetric about the vertical $z$ axis in a cylindrical coordinate system. Fig. 3(a) visualizes the drop profile along a sectional mid-plane passing through the drop centre. 
The
white
solid 1D curve in Fig. 3(a) represents the drop shape in the non-dimensional 
cartesian
coordinate space.

\subsection{Asymptotic analysis}
The asymptotic analysis in this section is based on the concepts developed in the context of solid particle impacts in gas flows \citep{koch1990kinetic,sundararajakumar1996non,how2021non}. We follow the analysis of How et al. \citep{how2021non} closely.
We begin with the asymptotic analysis of the air layer beneath the impacting drop. 
The asymptotic analysis analyzes the pressure and velocity field of the air layer just before the dimple starts to form.
Equation (3.1) could be rewritten for $r<<R_0$ using binomial expansion (region of interest is very close to the bottom part of the drop) as
\begin{equation}
    z(r) = R_0{\epsilon} + \frac{1}{2}\frac{r^2}{R_0}
\end{equation}
where ${\epsilon}=h_0/R_0$
is the asymptotic parameter characterizing the initial condition. Equation (3.2) depicts a parabolic (quadratic) profile approximating the bottom surface of the impacting drop. Normalizing the air layer thickness $z(r)$ with respect to minimum thickness of the air layer just before dimple formation $h_0={\epsilon}R_0$, equation (3.2) could be written in terms of a non dimensional thickness profile $\bar{z}$
\begin{equation}
    \bar{z}=\frac{z}{R_0{\epsilon}}=1+\frac{r^2}{2R_0^2{\epsilon}}
\end{equation}
where $\bar{z}=h(\bar{r})$ represents the non-dimensional height profile of the drop surface just before the dimple formation process.
Defining a non-dimensional radial coordinate $\bar{r}=r/(R_0{\epsilon}^{1/2})$ equation (3.3) could be written as
\begin{equation}
    \bar{z}=h(\bar{r})=\frac{z}{R_0{\epsilon}}=1+\frac{1}{2}\bar{r}^2
\end{equation}
The 
white
solid curve in Fig. 3(a) depicts equation (3.4) graphically along the sectional mid plane of the drop in cartesian coordinates $X-Z$ plane.  
The continuity equation in cylindrical axi-symmetric coordinates can be written as
\begin{equation}
    \frac{1}{r}\frac{\partial}{{\partial}r}\left(ru_r\right)+\frac{{\partial}u_z}{{\partial}z} = 0
    \label{e16}
\end{equation}

Using $z=\bar{z}R_0{\epsilon}$, ${r}=\bar{r}R_0{\epsilon}^{1/2}$, $u_z=\bar{u}_zV_0$ in equation (3.5) we have
\begin{equation}
    \frac{{\epsilon}^{1/2}}{V_0}\frac{1}{\bar{r}}\frac{\partial}{{\partial}\bar{r}}\left(\bar{r}{u}_r\right)+\frac{{\partial}\bar{u}_z}{{\partial}\bar{z}} = 0
\end{equation}
where $V_0$ is the impact velocity of the drop.
Defining $\bar{u}_r=u_r/(V_0/{\epsilon}^{1/2})$ the dimensionless continuity equation becomes 
\begin{equation}
  \frac{1}{\bar{r}}\frac{\partial}{{\partial}\bar{r}}\left(\bar{r}\bar{u}_r\right)+\frac{{\partial}\bar{u}_z}{{\partial}\bar{z}} = 0
\end{equation}

The radial momentum equation in cylindrical coordinates can be written as
\begin{equation}
    {\rho}_a\left(\frac{{\partial}{u}_r}{{\partial}t}+{u}_r\frac{{\partial}{u}_r}{{\partial}r}+{u}_z\frac{{\partial}{u}_r}{{\partial}z}\right)=-\frac{{\partial}{p}}{{\partial}r} + {\mu}_a\left(\frac{{\partial}^2{u}_r}{{\partial}{z}^2} + \frac{1}{r}\frac{{\partial}}{{\partial}r}\left(r\frac{{\partial}{u}_r}{{\partial}r}\right)\right)
\end{equation}
Using the scalings for $z=\bar{z}R_0{\epsilon}$, ${r}=\bar{r}R_0{\epsilon}^{1/2}$, $u_z=\bar{u}_zV_0$,
$u_r=\bar{u}_rV_0{\epsilon}^{-1/2}$, $t=\bar{t}R_0{\epsilon}/V_0$, $p=\bar{p}{\mu}_aV_0/(R_0{\epsilon}^2)$ in equation (3.8) the radial momentum equation becomes

\begin{equation}
    Re_{\epsilon}\left(\frac{{\partial}\bar{u}_r}{{\partial}\bar{t}}+\bar{u}_r\frac{{\partial}\bar{u}_r}{{\partial}\bar{r}}+\bar{u}_z\frac{{\partial}\bar{u}_r}{{\partial}\bar{z}}\right)=-\frac{{\partial}\bar{p}}{{\partial}\bar{r}} + \left(\frac{{\partial}^2\bar{u}_r}{{\partial}\bar{z}^2} + \frac{\epsilon}{\bar{r}}\frac{{\partial}}{{\partial}\bar{r}}\left(\bar{r}\frac{{\partial}\bar{u}_r}{{\partial}\bar{r}}\right)\right)
\end{equation}
where $Re_{\epsilon}={\rho}R_0V_0{\epsilon}/{\mu}_a=Re{\:}{\epsilon}$. 
$Re$ is the standard Reynolds number defined with respect to drop radius $R_0$, whereas $Re_{\epsilon}$ is the modified Reynolds number with respect to $R_0$. In the limit of ${\epsilon}{\rightarrow}0$ equation (3.9) reduces to
\begin{equation}
    \frac{{\partial}^2\bar{u}_r}{{\partial}\bar{z}^2}{=}\frac{{\partial}\bar{p}}{{\partial}\bar{r}}
\end{equation}
Equation (3.10) is the governing flow field equation near the leading edge of the drop. Equation (3.10) essentially describe continuum Stokesian flow in a quasi-steady framework \citep{sundararajakumar1996non,how2021non}. 
The $z$ momentum equation in cylindrical coordinates can be written as

\begin{equation}
    {\rho}_a\left(\frac{{\partial}{u}_z}{{\partial}t}+{u}_r\frac{{\partial}{u}_z}{{\partial}r}+{u}_z\frac{{\partial}{u}_z}{{\partial}z}\right)=-\frac{{\partial}{p}}{{\partial}z} + {\mu}_a\left(\frac{{\partial}^2{u}_z}{{\partial}{z}^2} + \frac{1}{r}\frac{{\partial}}{{\partial}r}\left(r\frac{{\partial}{u}_z}{{\partial}r}\right)\right)
\end{equation}

Using the scalings for $z=\bar{z}R_0{\epsilon}$, ${r}=\bar{r}R_0{\epsilon}^{1/2}$, $u_z=\bar{u}_zV_0$,
$u_r=\bar{u}_rV_0{\epsilon}^{-1/2}$, $t=\bar{t}R_0{\epsilon}/V_0$, $p=\bar{p}{\mu}_aV_0/(R_0{\epsilon}^2)$ in equation (3.11) the $z$ momentum equation becomes

\begin{equation}
    Re_{\epsilon}{\epsilon}\left(\frac{{\partial}\bar{u}_z}{{\partial}\bar{t}}+\bar{u}_r\frac{{\partial}\bar{u}_z}{{\partial}\bar{r}}+\bar{u}_z\frac{{\partial}\bar{u}_z}{{\partial}\bar{z}}\right)=-\frac{{\partial}\bar{p}}{{\partial}\bar{z}} + {\epsilon}\left(\frac{{\partial}^2\bar{u}_z}{{\partial}\bar{z}^2} + \frac{\epsilon}{\bar{r}}\frac{{\partial}}{{\partial}\bar{r}}\left(\bar{r}\frac{{\partial}\bar{u}_z}{{\partial}\bar{r}}\right)\right)
\end{equation}
In the limit of ${\epsilon}{\rightarrow}0$ equation (3.12) reduces to
\begin{equation}
     \frac{{\partial}\bar{p}}{{\partial}\bar{z}}=0
\end{equation}

Integrating equation (3.10) with respect to $\bar{z}$ twice we have
\begin{equation}
    {\bar{u}}_r(\bar{r},\bar{z}) = \frac{{\partial}\bar{p}}{{\partial}\bar{r}}\bar{{z}}^2 + C_1(\bar{r})\bar{z} + C_2(\bar{r})
\end{equation}
The quantities $C_1$ and $C_2$ are evaluated from the respective boundary conditions.
The continuum boundary conditions for the velocity field components are
\begin{equation}
    {\bar{u}}_r={\bar{u}}_z=0{\:};{\:}at{\:}\bar{z}=0
\end{equation}

\begin{equation}
    {\bar{u}}_r=0;{\:}{\bar{u}}_z=-1{\:};{\:}at{\:}\bar{z}=h(\bar{r})
\end{equation}

Evaluating $C_1$ and $C_2$ from equation (3.15) and equation (3.16); equation (3.14) for the radial velocity component reduces to
\begin{equation}
    {\bar{u}}_r = \frac{{\partial}\bar{p}}{{\partial}\bar{r}}\left(\frac{{\bar{z}}^2}{2} - \frac{\bar{z}h(\bar{r})}{2}\right)
\end{equation}
The pressure gradient term ${{\partial}\bar{p}}/{{\partial}\bar{r}}$ is still unknown at this stage and is computed from the continuity equation (3.7). Integrating equation (3.7) with respect $\bar{z}$ we have
\begin{equation}
    \int_0^{h(\bar{r})}\frac{1}{\bar{r}}\frac{{\partial}\bar{r}{\bar{u}}_r}{\partial{\bar{r}}}d\bar{z}+ \int_0^{h(\bar{r})}\frac{{\partial}{\bar{u}}_z}{{\partial}\bar{z}}d\bar{z}=0
\end{equation}
Using the limits in the second integral we have
\begin{equation}
    \int_0^{h(\bar{r})}\frac{1}{\bar{r}}\frac{{\partial}\bar{r}{\bar{u}}_r}{\partial{\bar{r}}}d\bar{z}+ {\bar{u}}_z|_{\bar{z}=h(\bar{r})}-{\bar{u}}_z|_{\bar{z}=0}= 0
\end{equation}
Applying the boundary conditions from equation (3.15) and (3.16) in equation (3.19), we have
\begin{equation}
    \int_0^{h(\bar{r})}\frac{1}{\bar{r}}\frac{{\partial}\bar{r}{\bar{u}}_r}{\partial{\bar{r}}}d\bar{z}+ (-1 - 0) = 0
\end{equation}

\begin{equation}
     \int_0^{h(\bar{r})}\frac{1}{\bar{r}}\frac{{\partial}\bar{r}{\bar{u}}_r}{\partial{\bar{r}}}d\bar{z} = 1
\end{equation}

Substituting equation (3.17) in equation (3.21) we have

\begin{equation}
    \frac{1}{\bar{r}}\frac{\partial}{\partial\bar{r}}\left[\bar{r}\int_0^{h(\bar{r})}\frac{{\partial}\bar{p}}{{\partial}\bar{r}}\left(\frac{{\bar{z}}^2}{2}-\frac{{\bar{z}h(\bar{r})}}{2}\right)d\bar{z}\right]=1
\end{equation}

\begin{equation}
    \frac{{\partial}\bar{p}}{{\partial}\bar{r}} = -\frac{6\bar{r}}{{h^3(\bar{r})}} + \frac{D_1}{\bar{r}{h^3(\bar{r})}}
\end{equation}

At $\bar{r}=0$, ${{\partial}\bar{p}}/{{\partial}\bar{r}}=0$ (symmetry boundary condition). Therefore $D_1=0$

\begin{equation}
    \frac{{\partial}\bar{p}}{{\partial}\bar{r}} = -\frac{6\bar{r}}{{h^3(\bar{r})}} 
\end{equation}

Substituting the radial pressure gradient from equation (3.24) in equation (3.17), the radial velocity could be written as
\begin{equation}
     {\bar{u}}_r = -\frac{6\bar{r}}{{h^3(\bar{r})}} \left(\frac{{\bar{z}}^2}{2} - \frac{\bar{z}h(\bar{r})}{2}\right)
\end{equation}

Using the undeformed asymptotic drop shape profile $h(\bar{r})=1+{\bar{r}}^2/2$ and integrating equation (3.24) 
we have

\begin{equation}
    \bar{p}(\bar{r}) = \frac{3}{(1+{\bar{r}}^2/2)^2}
\end{equation}
\begin{figure*}
    \centering
    \includegraphics[scale=0.8]{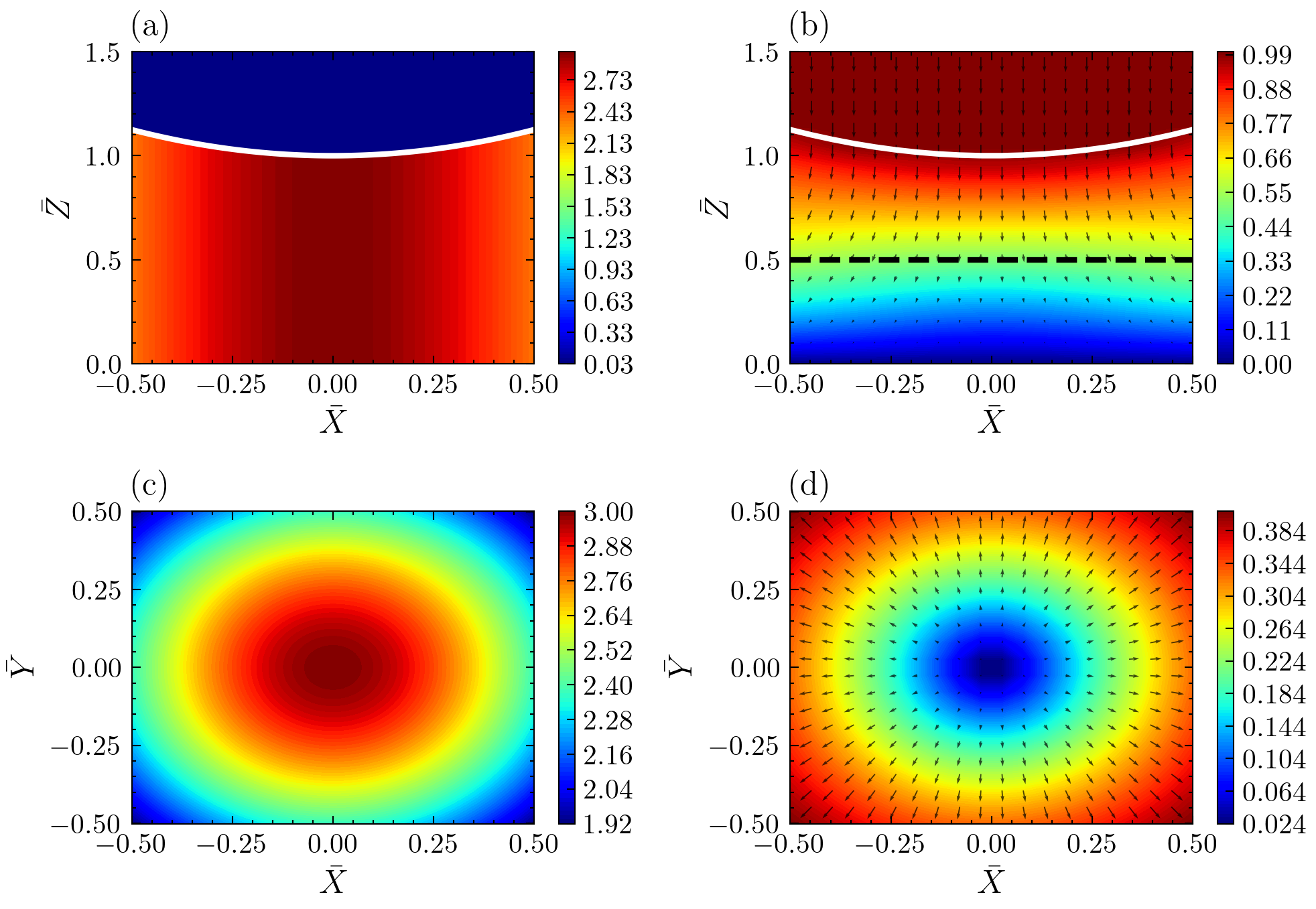}
    \caption{Asymptotic pressure and velocity fields just before the dimple formation. (a) Non dimensional pressure
    field in $\bar{X}$, $\bar{Z}$ plane. {The color bar represents non dimensional pressure field.}
    The white line depicts the air-water drop interface.
    (b) Non dimensional velocity field in $\bar{X}$, $\bar{Z}$ plane. {The color bar represents non dimensional velocity field.}(c) Pressure field in $\bar{X}$, $\bar{Y}$ plane. {The color bar represents non dimensional pressure field.}(d) Velocity field in $\bar{X}$, $\bar{Y}$ plane at $\bar{Z}=0.5$. The color bar represents non dimensional velocity field.}
    \label{Figure3}
    \color{black}
\end{figure*}
Further, substituting the dimple profile in equation (3.25) the radial velocity becomes
\begin{equation}
     {\bar{u}}_r = -\frac{6\bar{r}}{{(1+{\bar{r}}^2/2)}^{3}} \left(\frac{{\bar{z}}^2}{2} - \frac{\bar{z}}{2}{(1+{\bar{r}}^2/2)}\right)
\end{equation}

Fig. 3 depicts the asymptotic normalized pressure and velocity fields in the air layer beneath the drop just prior to dimple formation. Fig. 3(a) represents the axisymmetric pressure field in the $\bar{X}-\bar{Z}$ cartesian plane according to equation (3.26). The respective cartesian coordinates $x$, $y$, $z$ are expressed as non-dimensional cartesian coordinates using the scales ${x}=\bar{X}R_0{\epsilon}^{1/2}$, ${y}=\bar{Y}R_0{\epsilon}^{1/2}$, $z=\bar{Z}R_0{\epsilon}$. We observe from equation (3.26) that the non-dimensional pressure is maximum at the origin of coordinate system ($\bar{p}(\bar{r}=0)=3$).
The 
white
curve shows the asymptotic air-water drop interface prior to dimple formation begins according to equation (3.4). 
The actual experimental value of ${\epsilon}$ observed was ${\epsilon}=1.146{\times}10^{-3}$.
Fig. 3(b) represents the asymptotic velocity field data in the $\bar{X}$-$\bar{Z}$ plane using equation (3.27) and the linear approximation of the z-component of the velocity field $\bar{u}_z$ satisfying the boundary condition given by equation (3.15), (3.16) owing to the very small value of ${\epsilon}$ prior to dimple formation. The black dotted line represents $\bar{Z}=0.5$ horizontal plane parallel to $\bar{X}-\bar{Y}$ plane. Fig. 3(c) represents the axisymmetric pressure field in the $\bar{X}-\bar{Y}$ plane for $0<\bar{Z}<1$. Fig. 3(d) represents the axisymmetric velocity field in a plane given by $\bar{Z}=0.5$ parallel to the $\bar{X}-\bar{Y}$ plane according to equation (3.27). The velocity field denotes radially outward air flow from the center representing the air draining out due to the impacting drop.
The dimple starts to form when the air pressure at the center of the dimple becomes greater than the capillary pressure.
\begin{equation}
    p|_{\bar{r}=0}>p_{cap}
\end{equation}
The non dimensional pressure at the dimple centre is $\bar{p}(\bar{r}=0)=3$ from equation (3.26). Using the scale transformation between $p$ and $\bar{p}$ i.e., $p=\bar{p}{\mu}_aV_0/(R_0{\epsilon}^2)$ we have
\begin{equation}
    \frac{3{\mu}_aV_0}{R_0{\epsilon}^2} > \frac{2{\sigma}}{R_0}
\end{equation}
On simplifing further and solving the inequality for ${\epsilon}$ we have
\begin{equation}
    {\epsilon}<\sqrt{\frac{3{\mu}_aV_0}{2{\sigma}_{aw}}}
\end{equation}
Therefore the critical value of ${\epsilon}$ before central dimple formation begins is 
\begin{equation}
    {\epsilon}_{crit} = \sqrt{\frac{3}{2}}Ca^{1/2}
\end{equation}
where $Ca={\mu}_aV_0/{\sigma}_{aw}$ is the capillary number.

\begin{figure*}
    \centering
    \includegraphics[scale=0.5]{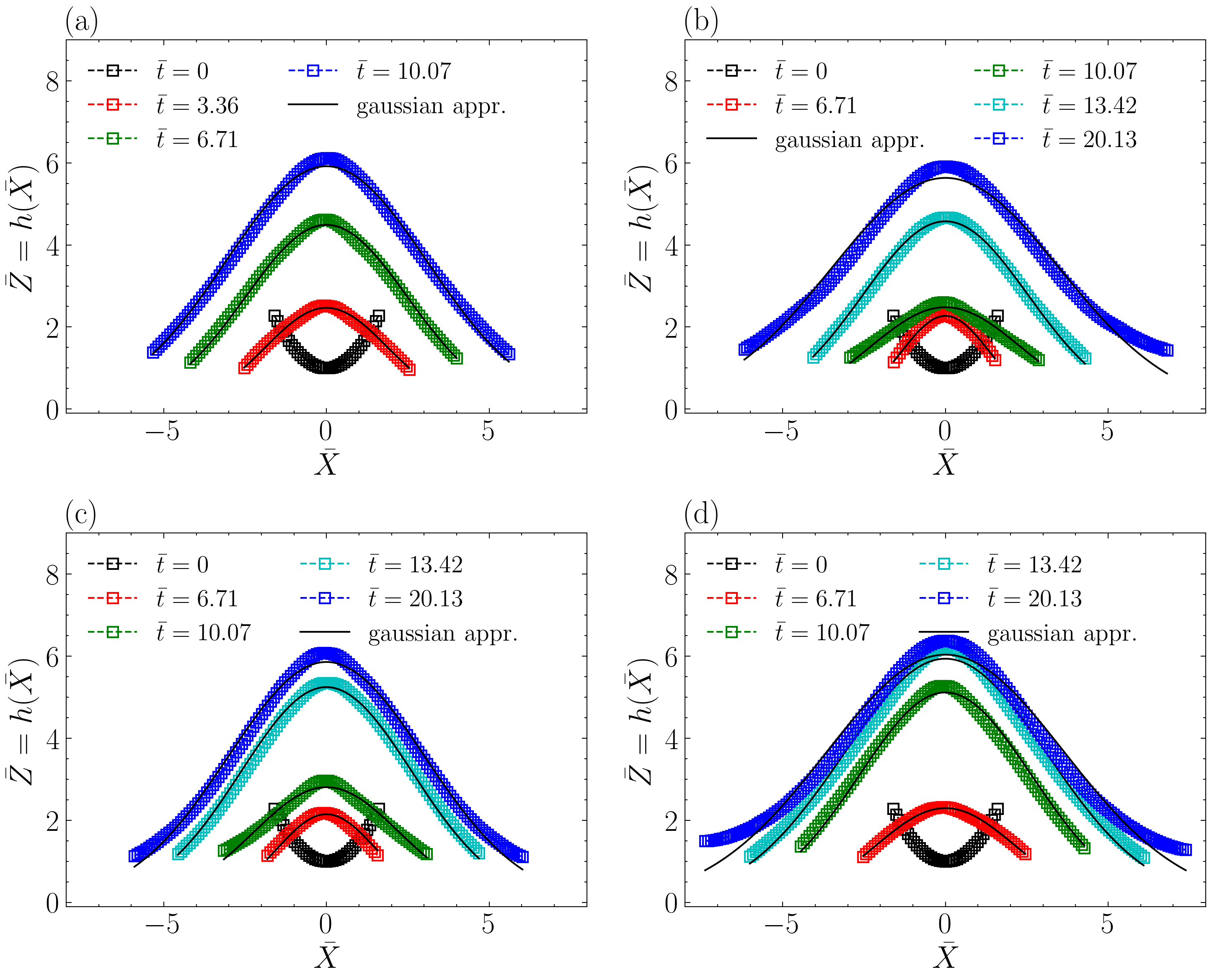}
    \caption{
    The evolution and the formation of the central dimple depicted by the variation of the 1D height profile as a function of time for various surface temperatures (a) $T_s=300$K (refer to supplementary Movie 1), (b) $T_s=353$K (refer to supplementary Movie 2), (c) $T_s=423$K (refer to supplementary Movie 3), (d) $T_s=473$K (refer to supplementary Movie 4). The horizontal and vertical axis $\bar{X}$, $\bar{Z}$ represents the non-dimensional cartesian coordinates.
    }
    \label{Figure4}
\end{figure*}

The actual central dimple thickness profile evolution was measured from high speed interferometric imaging as shown in Fig.4. Fig. 4(a), 4(b), 4(c) and 4(d) represents the central dimple evolution with time for substrate temperatures of  
$300$K, $353$K, $423$K and $473$K
respectively. The time coordinate has been renormalized appropriately for non-dimensional purposes as $\bar{t}=tV_0/(R_0{\epsilon})$. $\bar{t}=0$ is referenced with respect to the closest approach of the drop to the substrate prior to dimple formation, i.e. the time instant when ${\epsilon}=1.146{\times}10^{-3}$.  The shape of the drop close to the origin of coordinates is concave up shape represented by equation (3.4) locally (quadratic profile). Since at $\bar{t}=0$, ${\epsilon}<{\epsilon}_{crit}$, the pressure below the drop is greater than the capillary pressure (refer to equation 3.28-3.31). The excess pressure causes radial air flow outwards as depicted in Fig. 3(b) and 3(d). In addition to the flow as a response to the increase in air pressure, the air-water interface deforms and its curvature changes as depicted in Fig. 4 at different time instants labelled by $\bar{t}$.  We discover that the central air dimple evolution profiles can be approximated by a {G}aussian curve(refer to black curves in Fig.4) of the form
\begin{equation}
    h(\bar{r})=ae^{-\bar{r}^2/2{\sigma}^2}
\end{equation}

where $a$ characterizes the dimple thickness at the origin and ${\sigma}$ characterizes the spread of the profile (the standard deviation in terms of Gaussian), $\bar{r}$ is the radial coordinate.
\begin{figure*}
    \centering
    \includegraphics[scale=0.5]{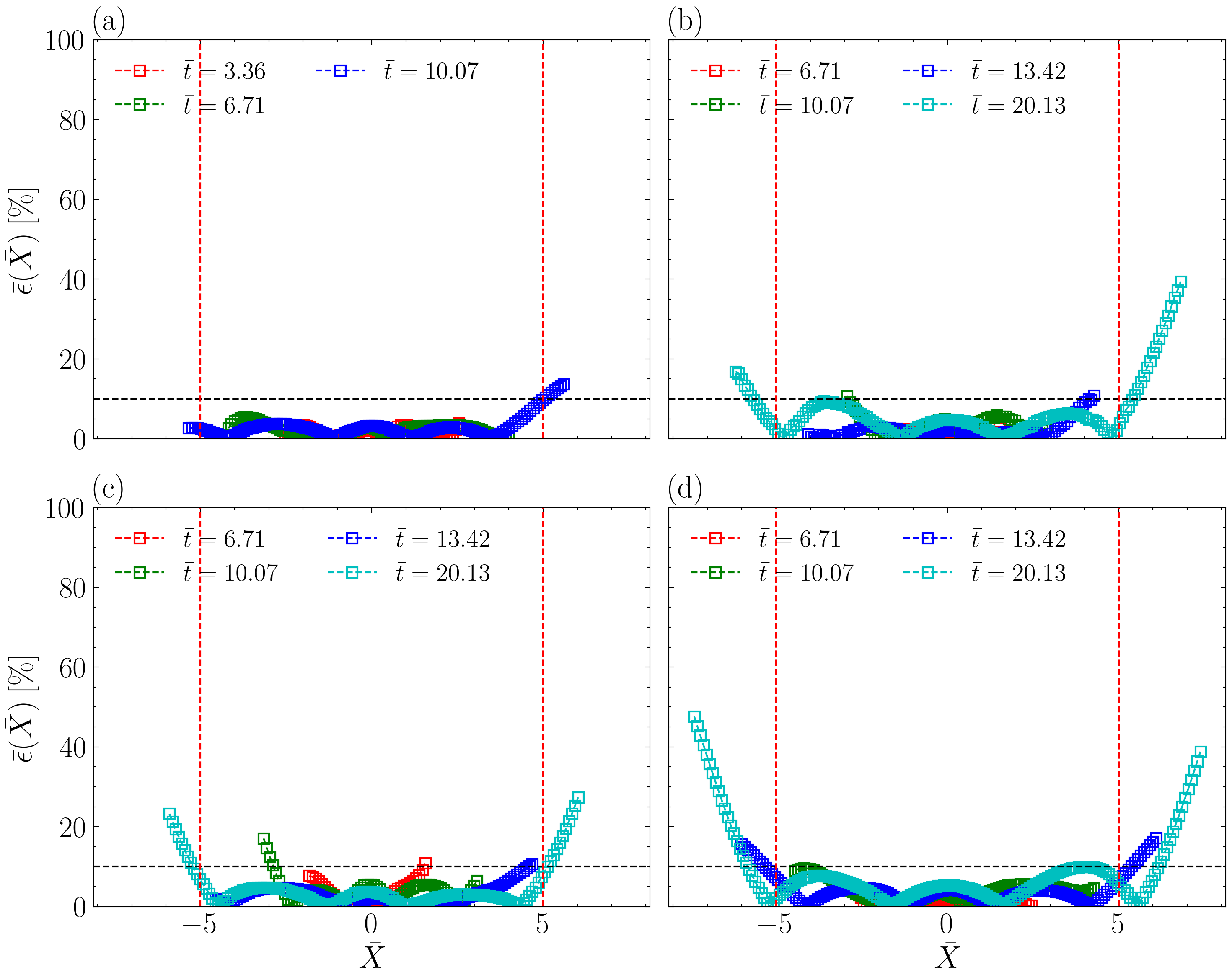}
    \caption{
    {
    The error percentage $\bar{\epsilon}(\bar{X})$ associated with approximating the central air dimple profile with the Gaussian approximation using equation (3.32) plotted as a function of non-dimensional $\bar{X}$ coordinate and non dimensional time $\bar{t}$ as a parameter for various surface temperatures $T_s$. The red vertical dotted line represents the coordinates $\bar{X}=-5$ and $\bar{X}=-5$. The black horizontal dotted line represents $\bar{\epsilon}=10\%$. 
    (a) $T_s=300$K, (b) $T_s=353$K, (c) $T_s=423$K, (d) $T_s=473$K. The horizontal and vertical axis $\bar{X}$ represents the non-dimensional cartesian coordinates.
    }}
    \label{Figure5}
\end{figure*}

The black curves in Fig.4 depicts the Gaussian approximation to the central dimple profile evolution at different time instants.
The Gaussian approximation is a good estimation 
of the the central air dimple height profile 
as can been seen from Fig. 5 for all substrate temperature.
Fig. 5 depicts the error percentage $\bar{\epsilon}$ plotted as a function of non dimensional x-coordinate $\bar{X}$ with non dimensional time $\bar{t}$ as a parameter for various surface temperature $T_s=300,{\:}353,{\:}423,{\:}473{\:}$K in Fig. 5(a), 5(b), 5(c), 5(d) respectively. $\bar{\epsilon}$ is calculated as 
$\bar{\epsilon}(\bar{X})=(|h_{exp}(\bar{X})-h_{G}(\bar{X})|{\times}100)/h_{exp}(\bar{X})$, 
where $h_{exp}(\bar{X})$ 
and 
$h_{G}(\bar{X})$
is the air layer thickness and the Gaussian approximation of the air layer height profile at a particular $\bar{X}$ coordinate respectively.
In Fig.5, $\bar{X}=-5,5$ is depicted as vertical dashed red lines. The horizontal dashed black line represents a $10\%$ error. It could be observed that the average error percentage (averaged over $\bar{X}$) in approximating the central air dimple is less than $5\%$ for $\bar{X}$ in the range $[-5,5]$ for all substrate temperature except $T_s=473$K. The error percentage $\bar{\epsilon}$ increases for $|\bar{X}|>5$, i.e. in the peripheral air disc region. Further, the author also wants to highlight that the error between the Gaussian fit and the experimental value increases with increase in temperature as can be seen for $T_s=473$K. However, the average error for $T_s=473$K shown in Fig. 5(d) is still smaller than $10\%$. This higher percentage error is due to the flattening nature of the dimple profile at higher temperatures caused due to combined effect of the weight of the droplet and the relatively stable air cushioning effect. Since, the air film evolution time scale increases with temperature, gravity and hence weight of the droplet along with the relatively thicker and stable air cushion can therefore effect the shape of the dimple.\\\\

Substituting equation (3.32) in equation (3.24) we have

\begin{equation}
    \frac{{\partial}\bar{p}}{{\partial}\bar{r}} = -\frac{6}{a^3}\bar{r}e^{3\bar{r}^2/2{\sigma}^2}
\end{equation}

Integrating with respect to $\bar{r}$ we have
\begin{equation}
    \bar{p}(\bar{r})=\bar{p}(0) - \int_0^{\bar{r}}\frac{6}{a^3}\bar{r}e^{3\bar{r}^2/2{\sigma}^2}d\bar{r}
\end{equation}

\begin{equation}
    \bar{p}(\bar{r})=\bar{p}(0) - \frac{2{\sigma}^2}{a^3}e^{3\bar{r}^2/2{\sigma}^2}
\end{equation}

Further on substituting equation (3.32) and (3.33) in equation (3.17) the radial velocity field becomes
\begin{equation}
     {\bar{u}}_r = -\frac{6}{a^3}\bar{r}e^{3\bar{r}^2/2{\sigma}^2}\left(\frac{\bar{z}^2}{2}-\frac{\bar{z}}{2}ae^{-\bar{r}^2/2{\sigma}^2}\right)
\end{equation}
\begin{figure*}
    \centering
    \includegraphics[scale=0.8]{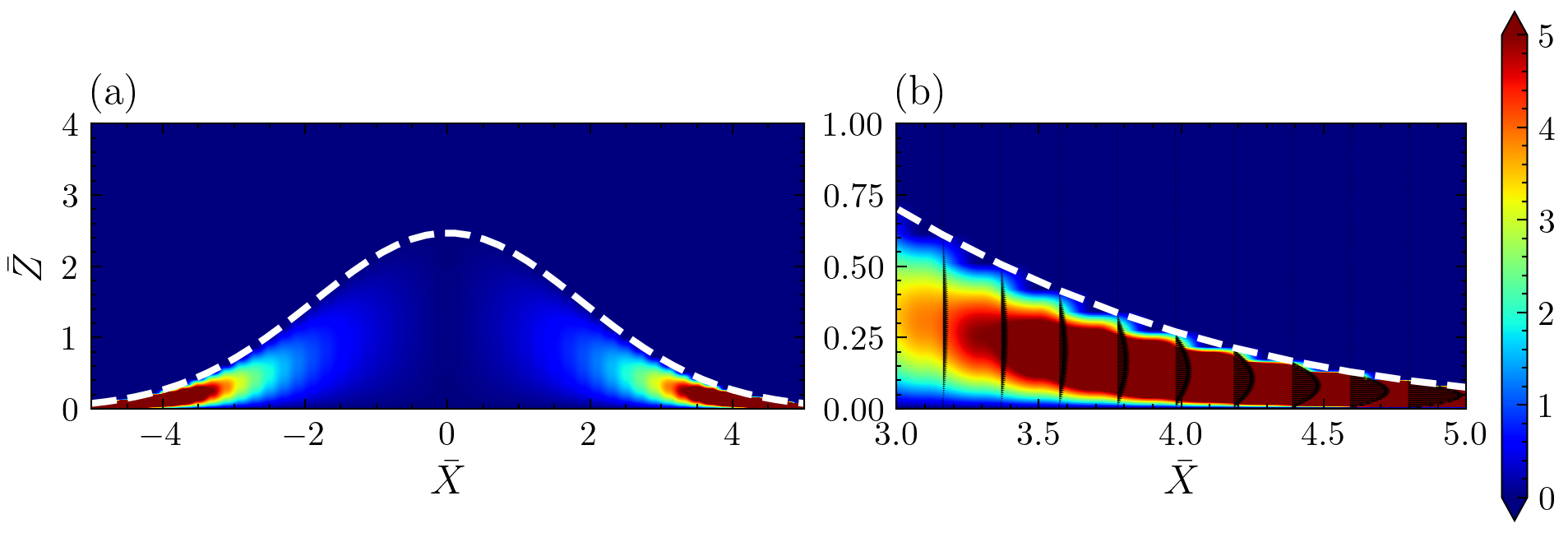}
    \caption{{The asymptotic radial velocity field depicted as a filled contour plot for substrate temperature
$T_s=300$K at $\bar{t}=3.36$ in the non dimensional cartesian coordinates (a) $\bar{X}=[-5,5]$, and $\bar{Y}=[0,4]$. (b) Zoomed in radial velocity field in the non dimensional cartesian coordinates $\bar{X}=[3,5]$, and $\bar{Y}=[0,1]$. The radial velocity profiles are also depicted for some discrete $\bar{X}$ using black arrows. The white dashed line depicts the air-water drop interface.} {The color bar represents non dimensional radial velocity field.}}
    \label{Figure6}
\end{figure*}

The radial velocity field variation with time is plotted as a contour field in 
Fig. 6, Fig. 7, and Fig. 8. Fig. 6 depicts the radial velocity field contours (Fig. 6(a)) and velocity profiles (Fig. 6(b)) in the $\bar{Z}$-$\bar{X}$ plane for $T_s=300$K
Figs. 7(a), 7(b) and 7(c)
represents the dimple evolution 
and its associated radial velocity contour fields
at three different time instants $\bar{t}=3.36$, $\bar{t}=6.71$ and $\bar{t}=10.07$
for $T_s=300$K. The 
white dotted
curve depicts the air-dimple interface evolution as a function of time according to equation (3.32) where $a$ and ${\sigma}$ are function of time. The time $\bar{t}=10.07$ corresponds to the instant where the dimple is at its maximum height and has attained a steady profile. It is interesting to note that the maximum radial velocity scale decreases as time progresses as can be observed from Fig. 7. This is in accordance with the continuity equation that as time progresses, the velocity profiles becomes elongated 
in the vertical direction due to increasing dimple height
and 
narrower in the horizontal direction denoting a decrease in velocity scale at a given fixed radial coordinate. Fig. 8
depicts the dynamic dimple evolution and the contour maps for impact on substrate at temperature $T_s=353K$. 
Fig. 8(a), 8(b), 8(c) and 8(d)
represents the various non-dimensional time instants represented by $\bar{t}=6.71$, $\bar{t}=10.07$, $\bar{t}=13.42$ and $\bar{t}=20.13$ respectively. 
Similar 
radial velocity
contours are plotted for other substrate temperatures in 
Fig. 2 ($423$K) and Fig. 3 ($473$K) of the supplementary document.
We can observe
that in all substrate temperature the non-dimensional radial velocity scales are of similar order of magnitude indicating the dynamics in the central dimple region is inherently independent of substrate temperature in the range considered in this work. 
The important idea that these plots reveal is the nature of temporal evolution of the radial velocity field. It should be noted that the asymptotic radial velocity and the its corresponding pressure field in the central dimple evolves as a quasi-steady field with the central dimple air-water interface (shown white dotted lines in the contour field plots) as the moving boundary.
The maximum central dimple thickness 
therefore
is weakly dependent on the substrate temperature. The region around the central dimple consists of the peripheral air disc where the air thickness is of the order of 
$100$nm. Small perturbations develops at the air-water interface
in the peripheral air disc region due to capillary and van der Waals interaction
(discussed in detail in a later section).
This
results in asymmetric air layer dewetting/rupture (or it could be called wetting of the drop with the substrate).
\begin{figure*}
    \centering
    \includegraphics[scale=0.8]{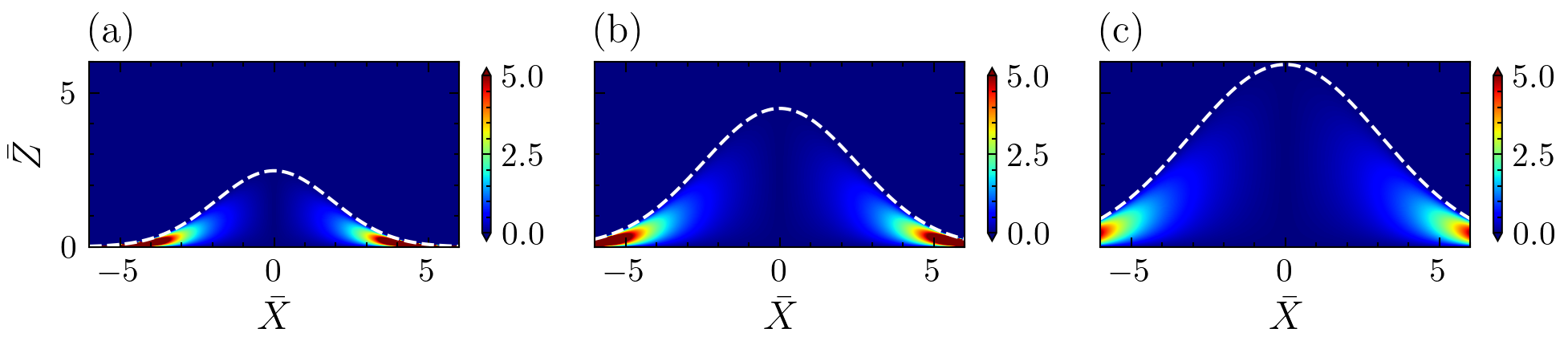}
    \caption{{The asymptotic radial velocity field evolution as a function of time in the $\bar{X}=[-5,5]$, $\bar{Z}=[0,5]$ cartesian plane depicted as a contour map for substrate temperature
$T_s=300$K at (a) $\bar{t}=3.36$, (b) $\bar{t}=6.71$, and (c) $\bar{t}=10.07$.} {The color bar represents non dimensional radial velocity field.}
    }
    \label{Figure7}
\end{figure*}

As the air between the drop and the substrate drains and thins out, structures appear in the peripheral air disc. The air thickness in the disc region reaches close to mean free path before first point of contact between the drop and the substrate. The first point of contact is designated as the air layer rupture point. The time it takes to form the rupture point is known as the rupture time scale and the radial distance between the rupture location and the origin of coordinates is known as the rupture radius.  
The rupture time scale increases with increase in substrate temperature as can be observed in the box-plot depicted in Fig. 9(a). We also observe correspondingly in Fig. 9(b) that the rupture radius increases with the substrate temperature $T_s$. The green horizontal line in the box plot denotes the median value of the rupture time and rupture radius. The rupture time and rupture radius are measured in 
${\mu}$s and ${\mu}$m
respectively. Corresponding to ${\epsilon}_{crit}$ from equation (3.31), the critical air layer thickness prior to dimple formation is given as 
\begin{equation}
    h_{crit} {\sim}  \sqrt{\frac{3}{2}}Ca^{1/2}R_0
\end{equation}
where $Ca={\mu}_aV_0/{\sigma}_{aw}$ is the capillary number and $R_0$ is the impacting drop radius.
The capillary number depends on the temperature due to surface tension 
and viscosity dependence of temperature as ${\sigma}_{aw}=a-bT_s$ and ${\mu}_a={\mu}_{ref}(T_s/T_{ref})^{\omega}$ \citep{valentini2023first} respectively. Here, ${\mu}_{ref}$ is the air viscosity at the reference temperature $T_{ref}$ which is $300K$ for the current study.
Incorporating the surface tension and
viscosity
dependence on temperature, the critical thickness from equation (3.37) becomes
\begin{equation}
    h_{crit}{\sim}\sqrt{\frac{3{\mu}_{ref}T^{\omega}_sV_0}{2{T^{\omega}_{ref}}(a-bT_s)}}R_0
\end{equation}
\begin{figure*}
    \centering
    \includegraphics[scale=0.8]{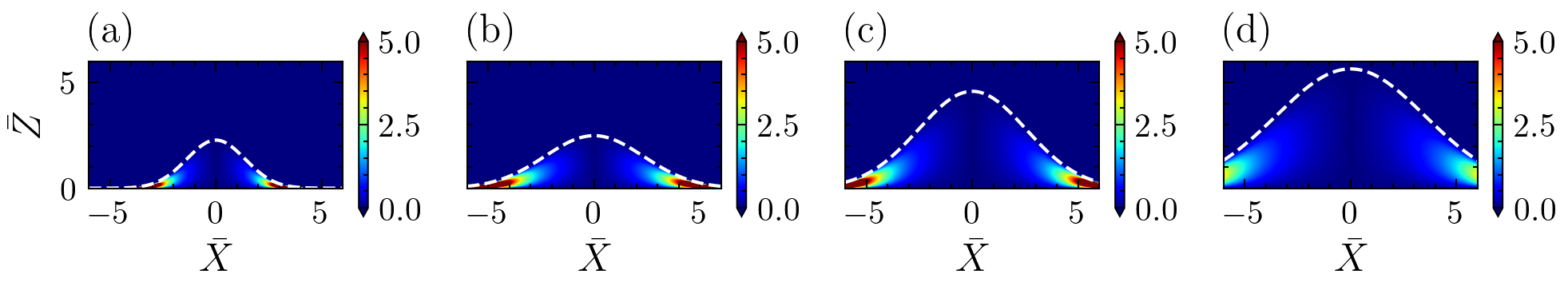}
    \caption{{The asymptotic radial velocity field evolution as a function of time in the $\bar{X}=[-5,5]$, $\bar{Z}=[0,5]$ cartesian plane depicted as a contour map for substrate temperature
 $T_s=353$K at (a) $\bar{t}=6.71$, (b) $\bar{t}=10.07$, (c) $\bar{t}=13.42$, and (d) $\bar{t}=20.13$.} {The color bar represents non dimensional radial velocity field.}}
    \label{Figure8}
\end{figure*}

\begin{figure*}
    \centering
    \includegraphics[scale=0.8]{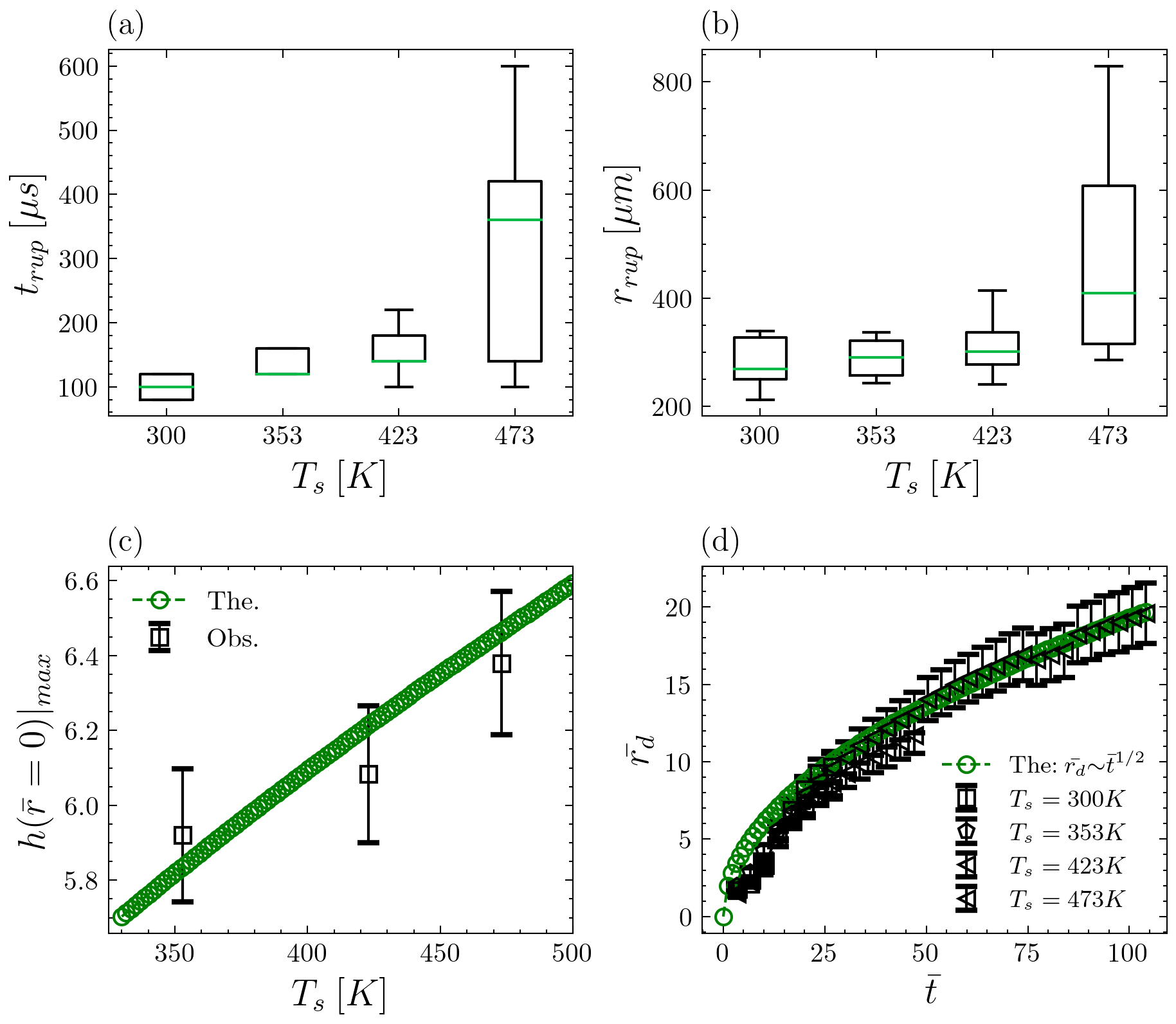}
    \caption{
    (a) The air layer rupture time as a function of substrate temperature.
(b) The air layer rupture radius as a function of substrate temperature.
(c) The central dimple maximum thickness as a function of substrate temperature.
(d) The radial air disc expansion as a function of time.
    }
    \label{Figure9}
\end{figure*}
In general the maximum central dimple thickness is proportional to the critical thickness given by equation (3.38) as the pressure at the center of the dimple that causes deformation of the drop interface is correlated to ${\epsilon}_{crit}$ and hence to $h_{crit}$.
Therefore, the maximum dimple thickness at $\bar{r}=0$ has a temperature dependence similar to equation (3.38) i.e. 
$h(\bar{r}=0)|_{max}{\sim}\sqrt{T^{\omega}_s}/\sqrt{a-bT_s}$.
Fig. 9(c) compares the experimental central dimple maximum thickness with the theoretical scaling of $h(\bar{r}=0)|_{max}$. The experimental values agrees with the theoretical scale within the experimental uncertainty.
The lubrication force prior to dimple formation is given by
\begin{equation}
    F_{lub} = \frac{2{\pi}{\mu}V_0R_0}{\epsilon}\int_0^{\bar{r}}\bar{p}(\bar{r})\bar{r}d\bar{r}
\end{equation}

Using the value of non-dimensional pressure from equation (3.26) in equation (3.39) we have 
\begin{equation}
    F_{lub} = \frac{6{\pi}{\mu}V_0R_0}{\epsilon}\int_0^{\bar{r}}\frac{\bar{r}d\bar{r}}{1+\bar{r}^2/2}=\frac{6{\pi}{\mu}V_0R_0}{\epsilon}\int_0^{\infty}\frac{\bar{r}d\bar{r}}{1+\bar{r}^2/2}=\frac{6{\pi}{\mu}V_0R_0}{\epsilon}
\end{equation}
\begin{figure*}
    \centering
    \includegraphics[scale=0.8]{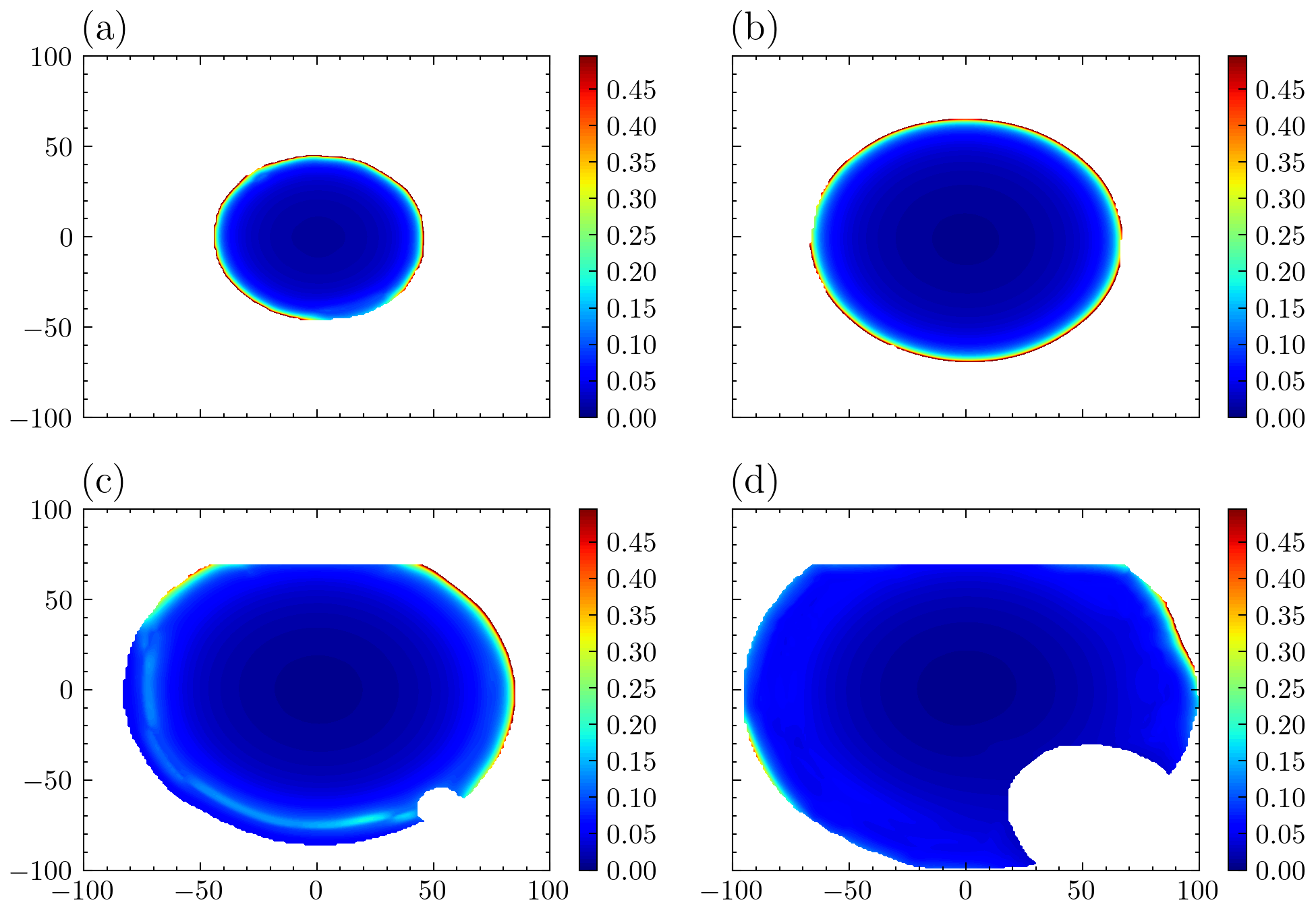}
    \caption{
    The 2D Knudsen field {$Kn={\lambda}/h$} evolution as a function of time for substrate temperature {${T}_s=300$K} at (a) $\bar{t}=6.71$, (b) $\bar{t}=10.07$, (c) $\bar{t}=13.42$, and (d) $\bar{t}=16.78$. {The color bar represents the Knudsen field.}}
    \label{Figure10}
\end{figure*}
The lubrication force varies inversely as ${\epsilon}$ and hence blows up in the limit ${\epsilon}{\rightarrow}0$. Physically for impact of solid objects the singularity is real and the singularity cutoff occurs through non-continuum effects. For the present case of impact of liquid drop, the singularity at the center of coordinates is avoided  due to deforming nature of the air water interface. i.e., for drop impacts on solids or liquids, the singularity is never reached at the center of coordinates due to dimple formation mechanisms discussed above.
For impact of drops on solids and liquids at moderate to low impact energies the system approaches the mathematical singularity in the peripheral air disc region. However, physically the singularity is never reached and is cut off by non-continuum 3d structures that occurs due to balance of capillary and van der Waal interactions.
Using ${\epsilon}=h_0/R_0$ in (3.40), the lubrication pressure prior to dimple formation becomes
\begin{equation}
    F_{lub}{\sim}\frac{6{\pi}{\mu}V_0R_0^2}{h_0}
\end{equation}
Applying Newtons second law provides us the dynamical equation of motion for the impacting drop,
\begin{equation}
    F_{mg} - F_b - F_{lub} - F_v= m\frac{dV_{cm}}{dt}
\end{equation}
where $F_{mg}$ is the weight of the drop, $F_b$ is the buoyancy force, $F_{lub}$ is the lubrication force, $F_v$ is the viscous drag, $m$ is the mass of the drop, $V_{CM}$ centre of mass velocity of the drop and $t$ is time. 
As the drop approaches the impacting surface just prior to dimple formation, the center of mass velocity of the drop attains approximately a constant value for the time duration under consideration, and hence $dV_{CM}/dt=0$. The non dimensional centre of mass $z$ coordinate $z_{CM}/R_0$ is plotted as a function of non dimensional time $t/{\tau}$ in Fig. 4 of the supplementary document. From the figure, we can observe that $V_{CM}$ is constant as ${z}_{CM}$ decreases linearly with time. Therefore, the centre of mass acceleration is zero ($dV_{CM}/dt=0$) just prior to dimple formation.
The Reynolds number is given as $Re={\rho}_aV_0R_0/{\mu}_a$, where ${\rho}_a$ is the air density, $V_0$ the impact velocity, $R_0$ is the drop radius, and ${\mu}_a$ is the air viscosity.
For, the current experimental condition using ${\rho}_a=1.293$kg/m$^3$, $V_0=0.25$m/s, $R_0=1.1$mm, ${\mu}_a=1.846{\times}10^{-5}$Pa.s, we have $Re=19.26{\sim}20$. In general, the drag on a spherical drop scales as $F_v{\sim}{\rho}_aV_0^2{\pi}R_0^2C_d$, where $C_d$ is the drag coefficient. For flow past spheres, $C_d{\sim}Re^{-1}$ for $Re<300$. Therefore the drag force on the drop scales as $F_v{\sim}{\pi}{\mu}_aR_0V_0$ (i.e. a linear drag law). Therefore, we use a linear drag model, the Stokes drag law. Further the deviation of the actual drag force from the Stokes drag is given by $C_dRe/24=F_v/6{\pi}{\mu}_aR_0V_0$. For $Re<200$, the factor $(C_dRe/24){\sim}\mathcal{O}(1)$ \cite{langmuir1946mathematical}. Therefore the drag force is given by $F_v{\sim}6{\pi}{\mu}_aR_0V_0$, i.e. the Stokes drag law. Using the Stokes drag as the scale of the viscous force, equation (3.42) scales as
\begin{equation}
    \frac{4}{3}{\pi}({\rho}_w-{\rho}_a)R_0^3g - \frac{6{\pi}{\mu}V_0R_0^2}{h_0} {\sim}  6{\pi}{\mu}R_0V_0
\end{equation}
On simplifying and solving for the scale of $h_0$ we have
\begin{equation}
    h_0 {\sim} \frac{9{\mu}_aV_0R_0^2}{2R_0^3g({\rho}_w - {\rho}_a) - 9{\mu}_aV_0R_0}
\end{equation}
Writing equation (3.44) in a non-dimensional manner we have
\begin{equation}
    {\epsilon}=\frac{h_0}{R_0} {\sim} \frac{9Ca}{2Bo - 9Ca}
\end{equation}
 where $Ca={\mu}_aV_0/{\sigma}_{aw}$, and $Bo = ({\rho}_w - {\rho}_a)R_0^2g/{\sigma}_{aw}$ is the Bond number based on the drop radius. The theoretical value of ${\epsilon}$ calculated using equation (3.45) is $1.25{\times}10^{-3}$ and agrees with the experimental value of $1.15{\times}10^{-3}$ within the experimental uncertainty.
\begin{figure*}
    \centering
    \includegraphics[scale=0.8]{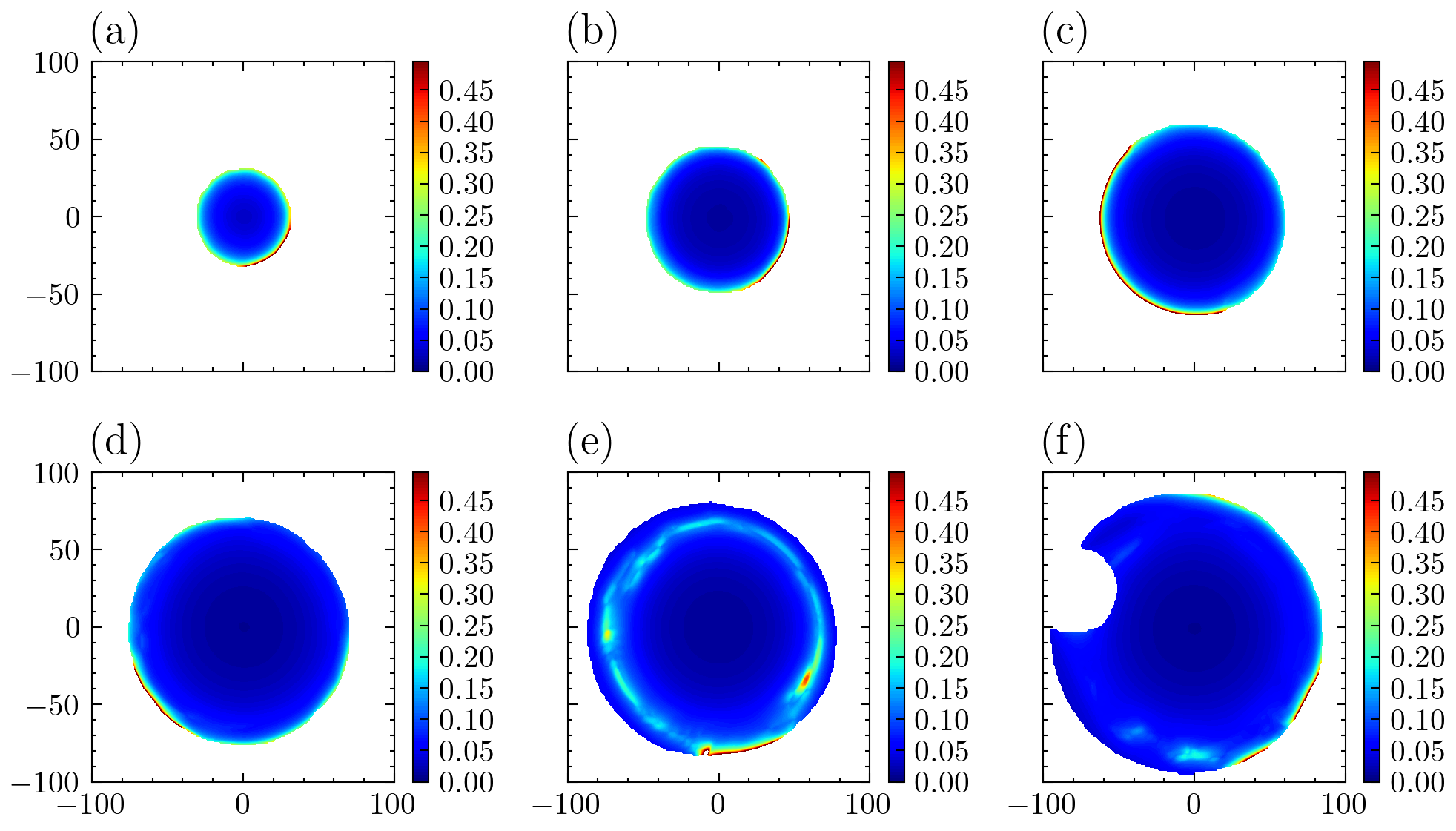}
    \caption{
    The 2D Knudsen field {$Kn={\lambda}/h$} evolution as a function of time for substrate temperature {${T}_s=353$K} at (a) $\bar{t}=6.71$, (b) $\bar{t}=10.07$, (c) $\bar{t}=13.42$, (d) $\bar{t}=16.78$, (e) $\bar{t}=20.13$, and (f) $\bar{t}=23.49$. {The color bar represents the Knudsen field.}}
    \label{Figure11}
\end{figure*}
\subsection{Estimating a scale for the dimple radius during dimple formation process}
The bottom surface of the drop and hence the air layer profile prior to dimple formation is given by (refer to equation (3.1))
\begin{equation}
    h(r) = h_0 + R_0\left(1-\sqrt{1-\left(\frac{r}{R_0}\right)^2}\right)
\end{equation}
The condition for the air profile height becoming equal to $h_{crit}$ at any particular radial coordinate $r_{dimp}$ is given by
\begin{equation}
    h_0 + R_0\left(1-\sqrt{1-\left(\frac{r_{dimp}}{R_0}\right)^2}\right) = h_{crit}
\end{equation}
The following inequalities are being satisfied in general impact scenarios. The condition for dimple to form in terms of air layer thickness is given by
\begin{equation}
    h_0<h_{crit}
\end{equation}
Condition given by equation (3.48) holds due to the following inequality that states, in a general impact scenario where the kinematic time scale of the interface is smaller than the capillary deformation time scale for the dimple formation, i.e.
\begin{equation}
    \frac{h_{crit}}{V_0}<\sqrt{\frac{{\rho}R^3}{{\sigma}_{aw}}}
\end{equation}
Rearranging equation (3.47) we have
\begin{equation}
    1-\sqrt{1-\left(\frac{r_{dimp}}{R_0}\right)^2} = \frac{h_{crit}-h_0}{R_0}
\end{equation}
Simplifying equation (3.50) we have
\begin{equation}
    \sqrt{1-\left(\frac{r_{dimp}}{R_0}\right)^2} = 1 - \left(\frac{h_{crit}-h_0}{R_0}\right)
\end{equation}
The dimple radius therefore becomes
\begin{equation}
    \frac{r_{dimp}}{R_0}=\sqrt{f(2-f)}
\end{equation}
where
\begin{equation}
    f = \frac{h_{crit}-h_0}{R_0}={\epsilon}_{crit}-{\epsilon}
\end{equation}
Note equation (3.52) predicts the approximate radius of dimple during its formation stage. Therefore equation (3.52) should provide a dimple scale that should be greater than the radial length scale $R_0{\epsilon}^{1/2}$ and smaller than the maximum dimple radius, i.e.
\begin{equation}
    R_0{\epsilon}^{1/2} < r_{dimp} < r_{dimp}|_{max}
\end{equation}
Using the drop radius $R_0$ and the value of ${\epsilon}$ and ${\epsilon}_{crit}$ from equations (3.45) and (3.27) respectively, the inequality provided in (3.54) is satisfied (the numerical translation of inequality (3.54) is 
$46{\mu}m<142{\mu}m<509{\mu}m$).
\begin{figure*}
    \centering
    \includegraphics[scale=0.8]{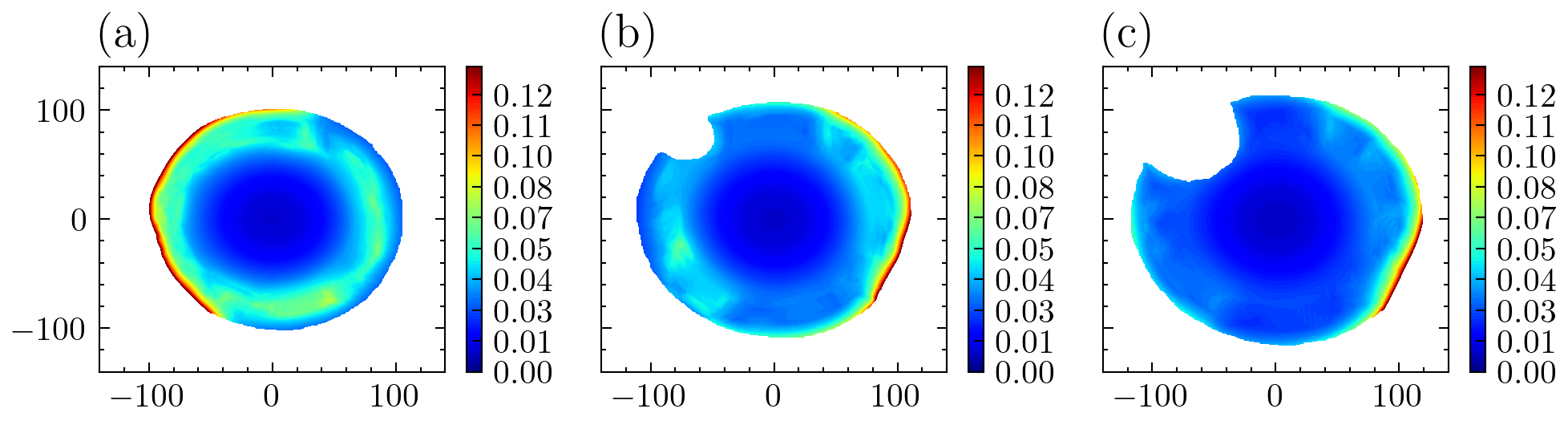}
    \caption{
    The 2D Knudsen field {$Kn={\lambda}/h$} evolution as a function of time for substrate temperature 
    {${T}_s=423$K at 
    (a) $\bar{t}=26.85$, (b) $\bar{t}=30.20$, and (c) $\bar{t}=33.56$
    }. {The color bar represents the Knudsen field.}
    }
    \label{Figure12}
\end{figure*}

\subsection{The expansion of the peripheral air disc}
Based on the Stokes approximation in the air layer region, the radial pressure gradient balances the viscous stresses and hence the radial momentum equation becomes
\begin{equation}
    \frac{{\partial}{p}}{{\partial}{r}}={\mu}_a\frac{{\partial}^2{u}_r}{{\partial}z^2}
\end{equation}
The pressure in the air layer is a superposition of various effects, however capillary is the dominant scale once the dimple has formed. Therefore using the scales ($r{\sim}r_d$, $z{\sim}h_0$, 
$p{\sim}2{\sigma}_{aw}/R_0$, $u_r{\sim}<V>$ where $<V>=dr_d/dt$ is the average expansion velocity of the air disc) in equation 
(3.55)
\begin{equation}
    \frac{1}{r_d}\frac{2{\sigma}_{aw}}{R_0}{\sim}{\mu}_a\frac{<V>}{{h_0}^2}=\frac{{\mu}_a}{{h_0}^2}\frac{dr_d}{dt}
\end{equation}

\begin{equation}
    r_d\frac{dr_d}{dt}{\sim}\frac{2{\sigma}_{aw}}{R_0}\frac{{h_0}^2}{{\mu}_a}
\end{equation}
\begin{figure*}
    \centering
    \includegraphics[scale=0.8]{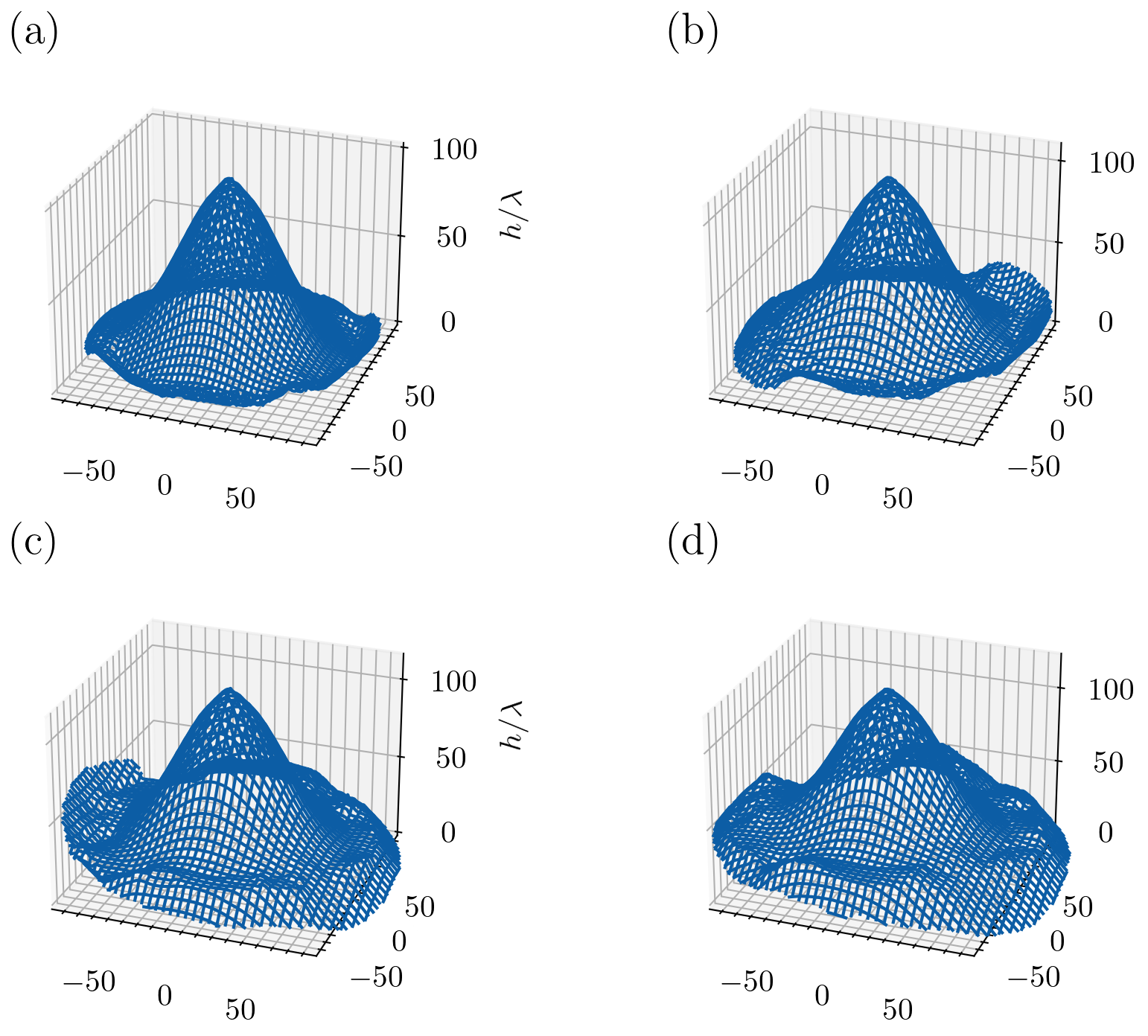}
    \caption{
    {
    The 3D non dimensional air layer thickness profile $h/{\lambda}$ as a function of time for substrate temperature ${T}_s=423$K at (a) $\bar{t}=23.49$, (b) $\bar{t}=26.85$, (c) $\bar{t}=30.20$, and (d) $\bar{t}=33.56$.
    }
    }
    \label{Figure13}
\end{figure*}
Integrating with respect to $t$ we have

\begin{equation}
    \int{r_ddr_d}{\sim}\int\frac{2{\sigma}_{aw}}{R_0}\frac{{h_0}^2}{{\mu}_a}dt
\end{equation}
The radial disc expansion as a function of time becomes
\begin{equation}
    r_d{\sim}2\sqrt{\frac{{\sigma}_{aw}{h_0}^2}{R_0{\mu}_a}}t^{1/2}
\end{equation}
Fig. 9(d) depicts the air disc expansion both theoretically and experimentally. The experimental and theoretical scales agree within the experimental uncertainities. Notice that from equation (3.59) we have $r_d{\sim}t^{1/2}$ (Fig. 9(d))
and the coefficient of $t^{1/2}$ predicted using equation (3.59) agrees with the experiments.

\subsection{Spatial length scales in the peripheral air disc}
The central air dimple is surrounded by a thin region that we refer to as the peripheral air disc. The peripheral air disc thickness is one order smaller than the central dimple thickness and approaches the mean free path of air.
The mean free path (${\lambda}$) of the air molecules is given by 
\begin{equation}
    {\lambda} = \frac{{\mu}_a}{p}\sqrt{\frac{{\pi}R_{s}T}{2}}
     \label{e10}
\end{equation}
where ${\mu}_a$ is the air viscosity, and $p$, $T$, $R_s$ denote air pressure, temperature and specific gas constant respectively. At standard room temperature of 
$300$K
and pressure of
$10^5{
\:}$Pa
the mean free path is approximately 
$67$nm.
The ratio of the mean free path of the air to local air layer thickness profile represents a 2D Knudsen field that we use to characterize the air layer beneath the impacting drop. The Knudsen field is defined as
\begin{equation}
    Kn=\frac{{\lambda}}{h}
\end{equation}
where $h$ represents the two dimensional air layer thickness profile. Note that the reciprocal of the Knudsen field represents a non-dimensional height profile measure. 
Figs. 10, 11, 12, 14 depicts the Knudsen field plotted for substrate temperatures of 
$300$K, $353$K, $423$K, $473$K respectively
at various non dimensional time instants. Fig. 5 and Fig. 6 of the supplementary document shows the Knudsen field variation for a larger temporal interval for $T_s=423$K and $T_s=473$K respectively.
Comparing the Knudsen field, we can observe that the structures in the peripheral disc regions are distinct. A unique circular pattern is observed for all the temperature cases in the peripheral disc region where $Kn{\sim}{\mathcal{O}(10^{-1})}$ (refer to Figs. 10(c), 11(e), 12(a), 14(d)). Subjected to very high values of Knudsen number in the peripheral air disc region, the dynamics is governed by an interaction of continuum and non-continuum effects. The flow field in the peripheral disc region lies in the slip flow and transition flow regime in contrast to the central air dimple. Notice the the evolving nature of the structures in the peripheral air disc denoting a time varying unsteady field  
(Fig. 13). Fig. 13 shows the 3D visualization of a typical air layer thickness profile as a function of non dimensional time for substrate temperature $T_s=423$K. The evolving perturbative structures (air layer height fluctuations) in the peripheral air disc region are clearly visible.
This is in contrast to the central dimple which is essentially quasi-steady
and relatively smooth. The structure in the peripheral air disc and the relatively smooth profile could be understood in terms of capillary flows as discussed below.
As the air beneath the impacting drop drains out in the radial direction, the air-water interface at the bottom of the drop is subjected to capillary waves.
As the air drains out below the impacting drop, the rise in capillary pressure causes the formation of the dimple. The draining air along with high curvatures caused during during dimple formation produces capillary waves at the bottom air-water interface of the drop \citep{mandre2009precursors}. Capillary waves could be sustained if
the capillary wavelength given by ${\lambda}_c$ is smaller than equal to the drop length scale $R_0$, i.e.,
\begin{equation}
    {\lambda}_c=\frac{{\sigma}_{aw}}{{\rho}_lV^2}{\leq}R_0
\end{equation}
where ${\rho}_l$ is the density of the liquid, ${\sigma}_{aw}$ is the air-water surface tension, and $V$ is the 
air velocity between the drop and the substrate. Equation (3.62) is a necessary but not a sufficient condition for capillary waves to form.
For the impact condition in this work, the 
inequality
is satisfied and hence the interface is subjected to capillary perturbations. 
From Figs. 10-14, we observe that the Knudsen number before the first point of contact in certain locations of the peripheral air disc is one order larger than the mean free path ($Kn{\sim}0.1$). Owing to the very small gap between the substrate and drop, the pressure in the gap region is subject to disjoining pressure that originates due to molecular interactions between surfaces. Derjaguin \citep{derjaguin1978question} showed that the pressure field in such small gaps becomes anisotropic in comparison to the large scale isotropic pressure. The balance between disjoining and capillary pressure, establishes the existence of new length scales for short time scales.

\begin{figure*}
    \centering
    \includegraphics[scale=0.7]{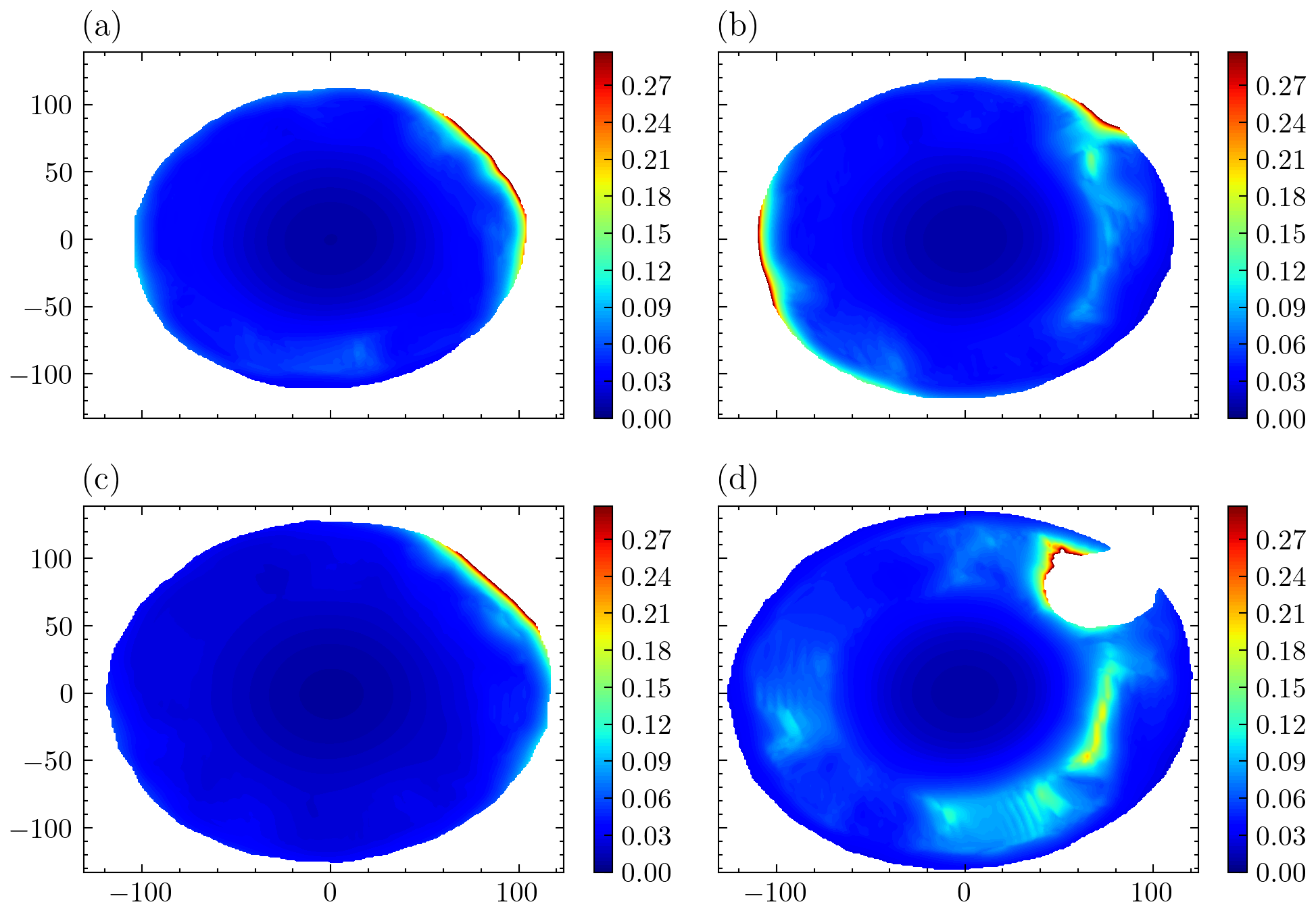}
    \caption{
    The 2D Knudsen field {$Kn={\lambda}/h$} evolution as a function of time for substrate temperature  
    { ${T}_s=473$K at
    (a) $\bar{t}=23.49$, (b) $\bar{t}=26.85$, (c) $\bar{t}=30.20$, and (d) $\bar{t}=33.56$}. {The color bar represents the Knudsen field.}
    }
    \label{Figure14}
\end{figure*}

The molecular (disjoining pressure term) and capillary interaction balance becomes \citep{roy2022droplet}
\begin{equation}
    \frac{A}{{h_{pd}}^3}{\sim}\frac{{\sigma}_{aw}}{R_0}
\end{equation}
where $A$ is the Hamacker's constant 
($A{\sim}\mathcal{O}$($4.76{\times}10^{-20}$J)),
$h_{pd}$ is the thickness in the peripheral air disc region. Solving for $h_{pd}$ from equation (3.63) becomes
\begin{equation}
    h_{pd}{\sim}\left(\frac{AR_0}{{\sigma}_{aw}}\right)^{1/3}
\end{equation}
The wavelength corresponding to thin film instability is given by \citep{vrij1966possible,roy2022droplet}
\begin{equation}
    {\lambda}_{TF}{\sim}h^2\left(\frac{{\sigma}_{aw}}{A}\right)^{1/2}
\end{equation}
    where $h$ represents the air layer thickness in the peripheral air disc region. Substituting $h_{pd}$ from equation (3.64) in equation (3.65) we have
\begin{equation}
    {\lambda}_{TF}{\sim}\left(\frac{A{R_0}^4}{{\sigma}_{aw}}\right)^{2/3}\left(\frac{{\sigma}_{aw}}{A}\right)^{1/2}
\end{equation}
Simplifying equation (3.66) we get
\begin{equation}
    {\lambda}_{TF}{\sim}\left(\frac{A{R_0}^4}{{\sigma}_{aw}}\right)^{1/6}
\end{equation}

Fig. 7 in the supplementary document shows the length scale measurement process in the peripheral air disc. Fig. 7(a) in the supplementary document shows a typical Knudsen field for $T_s=300$K. The perturbative structures caused due to the combined effect of capillary effects, thin film instability and intermolecular forces can be observed by a distinct cyan color in the disc region. A typical black dotted line shown depicts the measurement pathway of the length scale. Fig. 7(b) in the supplementary document depicts the intensity profile along the black dotted line. The length scale is determined from the width of the peak near the maximum intensity value. The experimental length scale is approximately $12{\mu}$m 
(${\sim}\mathcal{O}$($10{\mu}$m))
with a standard deviation of $4{\mu}$m. 
The theoretical capillary wavelength ${\lambda}_c$ and the thin film instability length scale ${\lambda}_{TF}$ predicted from equation (3.63) and (3.67) respectively corroborates with the experimental measured scales in the peripheral disc region. The theoretical scale comes out to be of the order of ${\lambda}_c{\sim}8{\mu}$m and  
${{\lambda}_{TF}}{\sim}7{\mu}$m.
The experimental length scale of the order of $10{\mu}$m corroborates with the theoretical values of the capillary wavelength and the thin film wavelength within the experimental uncertainty. The central air dimple interface is smoother than
the peripheral air disc interface as could be seen in the various
Knudsen field contour plots. The relative smoothness of the interface could be understood based on the capillary wave characteristics given by equations (3.62). We can observe that the capillary wavelength scale ${\lambda}_c$ is inversely proportional to the air velocity field between the drop and the substrate. It could be observed from equation (3.36), that the average radial air velocity increases with radial coordinate. Therefore, for higher velocities observed in the peripheral air disc corresponds to lower wavelength capillary waves in comparison to smaller velocities and larger wavelength capillary waves near the central dimple region. The central air dimple is therefore subjected long wave capillary perturbations in contrast to short wave capillary perturbations in the peripheral air disc region. The short wave capillary perturbations along with the van der Waals interactions results in the formation of hole nucleation points that result in air layer dewetting.

For the impact Weber number of unity considered in this work ($We{\sim}1$), the first point of contact occurs through kink mode of contact as shown numerically by Chubynsky et al. For larger Weber number regime that we have observed in our previous work (for impact on liquid pools) we observe the first point of contact occurs through film mode of contact. Therefore, the structures we discover in the peripheral air disc for a variety of imapct events are essentially the experimental realizations of the kink and film mode of contact discussed numerically in some recent articles \citep{de2015air-1,chubynsky2020bouncing}. 

\section{Conclusion}
In conclusion, we study the air layer dynamics beneath an impacting drop on a heated substrate. We show that a Gaussian profile could represent the evolving central air dimple profile. The geometry of the dimple depends on the substrate temperature very weakly and is a function of the impact Weber number majorly. Further, we observe that the air layer rupture time scale and rupture radius increase with substrate temperature slowly. We characterize the entire air layer profile beneath the drop by a time-varying Knudsen field. We show that a unified treatment of continuum and non-continuum mechanics is required to comprehend the flow field and the local anisotropic structures observed in the peripheral air disc. The airflow dynamics in the central dimple region lies within the continuum Stokes regime, whereas the peripheral air disc falls within the non-continuum slip flow and transition regime owing to the large Knudsen number. However, the average air disc expansion length scale on an average could be explained based on the continuum Stokes approximation. We further show that the structures observed in the peripheral air disc provide the mechanism for the film and kink contact modes between the drop and the substrate. The structures in the peripheral disc region initiate asymmetric contact between the drop and the substrate. The structures in the peripheral air disc occur due to the asymptotic effects of van der Waals and capillary interactions. 
We hope that the current work will lead to 
future investigations at the intersection of classical statistical and continuum mechanics to probe the mechanisms and the associated spatiotemporal length scales in the peripheral air disc region in 
further
detail.

%\begin{figure}
%  \centerline{\includegraphics{Figure2.pdf}}% Images in 100% size
%  \caption{xxx}
%\label{Figure2}
%\end{figure}

\section*{Supplementary movie captions}
Movie 1: High speed interferometry visualization for drop impact on substrate temperature $T_s=300K$

Movie 2: High speed interferometry visualization for drop impact on substrate temperature $T_s=353K$

Movie 3: High speed interferometry visualization for drop impact on substrate temperature $T_s=423K$

Movie 4: High speed interferometry visualization for drop impact on substrate temperature $T_s=473K$
% \section*{Acknowledgement}

\section*{Declaration of Interests}
The authors declare no conflict of interest.
\clearpage
\bibliographystyle{jfm}
\bibliography{jfm-instructions}

\begin{thebibliography}{70}
\expandafter\ifx\csname natexlab\endcsname\relax\def\natexlab#1{#1}\fi
\def\au#1{#1} \def\ed#1{#1} \def\yr#1{#1}\def\at#1{#1}\def\jt#1{\textit{#1}}
  \def\bt#1{#1}\def\bvol#1{\textbf{#1}} \def\vol#1{#1} \def\pg#1{#1}
  \def\publ#1{#1}\def\arxiv#1{#1}\def\org#1{#1}\def\st#1{\textit{#1}}

\bibitem[Andrade {\em et~al.\/}(2013)Andrade, Skurtys \&
  Osorio]{andrade2013drop}
{\sc \au{Andrade, R}, \au{Skurtys, O} \& \au{Osorio, F}} \yr{2013}  \at{Drop
  impact behavior on food using spray coating: Fundamentals and applications}.
  \jt{Food research international}  \bvol{54}~(1),  \pg{397--405}.

\bibitem[Bourouiba(2021)]{bourouiba2021fluid}
{\sc \au{Bourouiba, Lydia}} \yr{2021}  \at{The fluid dynamics of disease
  transmission}.  \jt{Annual Review of Fluid Mechanics}  \bvol{53},
  \pg{473--508}.

\bibitem[Bouwhuis {\em et~al.\/}(2012)Bouwhuis, van~der Veen, Tran, Keij,
  Winkels, Peters, van~der Meer, Sun, Snoeijer \& Lohse]{bouwhuis2012maximal}
{\sc \au{Bouwhuis, Wilco}, \au{van~der Veen, Roeland~CA}, \au{Tran, Tuan},
  \au{Keij, Diederik~L}, \au{Winkels, Koen~G}, \au{Peters, Ivo~R}, \au{van~der
  Meer, Devaraj}, \au{Sun, Chao}, \au{Snoeijer, Jacco~H} \& \au{Lohse, Detlef}}
  \yr{2012}  \at{Maximal air bubble entrainment at liquid-drop impact}.
  \jt{Physical review letters}  \bvol{109}~(26),  \pg{264501}.

\bibitem[Breitenbach {\em et~al.\/}(2017)Breitenbach, Roisman \&
  Tropea]{breitenbach2017heat}
{\sc \au{Breitenbach, Jan}, \au{Roisman, Ilia~V} \& \au{Tropea, Cameron}}
  \yr{2017}  \at{Heat transfer in the film boiling regime: Single drop impact
  and spray cooling}.  \jt{International Journal of Heat and Mass Transfer}
  \bvol{110},  \pg{34--42}.

\bibitem[Breitenbach {\em et~al.\/}(2018)Breitenbach, Roisman \&
  Tropea]{breitenbach2018drop}
{\sc \au{Breitenbach, Jan}, \au{Roisman, Ilia~V} \& \au{Tropea, Cameron}}
  \yr{2018}  \at{From drop impact physics to spray cooling models: a critical
  review}.  \jt{Experiments in Fluids}  \bvol{59},  \pg{1--21}.

\bibitem[Chubynsky {\em et~al.\/}(2020)Chubynsky, Belousov, Lockerby \&
  Sprittles]{chubynsky2020bouncing}
{\sc \au{Chubynsky, Mykyta~V}, \au{Belousov, Kirill~I}, \au{Lockerby, Duncan~A}
  \& \au{Sprittles, James~E}} \yr{2020}  \at{Bouncing off the walls: the
  influence of gas-kinetic and van der waals effects in drop impact}.
  \jt{Physical Review Letters}  \bvol{124}~(8),  \pg{084501}.

\bibitem[Daniel {\em et~al.\/}(2017)Daniel, Timonen, Li, Velling \&
  Aizenberg]{daniel2017oleoplaning}
{\sc \au{Daniel, Dan}, \au{Timonen, Jaakko~VI}, \au{Li, Ruoping}, \au{Velling,
  Seneca~J} \& \au{Aizenberg, Joanna}} \yr{2017}  \at{Oleoplaning droplets on
  lubricated surfaces}.  \jt{Nature Physics}  \bvol{13}~(10),  \pg{1020--1025}.

\bibitem[Davis(1984)]{davis1984rate}
{\sc \au{Davis, Robert~H}} \yr{1984}  \at{The rate of coagulation of a dilute
  polydisperse system of sedimenting spheres}.  \jt{Journal of Fluid Mechanics}
   \bvol{145},  \pg{179--199}.

\bibitem[De~Ruiter {\em et~al.\/}(2015)De~Ruiter, Lagraauw, Van Den~Ende \&
  Mugele]{de2015wettability}
{\sc \au{De~Ruiter, Jolet}, \au{Lagraauw, Rudy}, \au{Van Den~Ende, Dirk} \&
  \au{Mugele, Frieder}} \yr{2015}  \at{Wettability-independent bouncing on flat
  surfaces mediated by thin air films}.  \jt{Nature physics}  \bvol{11}~(1),
  \pg{48--53}.

\bibitem[Derjaguin \& Churaev(1978)]{derjaguin1978question}
{\sc \au{Derjaguin, BV} \& \au{Churaev, NV}} \yr{1978}  \at{On the question of
  determining the concept of disjoining pressure and its role in the
  equilibrium and flow of thin films}.  \jt{Journal of Colloid and Interface
  Science}  \bvol{66}~(3),  \pg{389--398}.

\bibitem[Dowling \& Dowling(2013)]{dowling2013scaling}
{\sc \au{Dowling, David~R} \& \au{Dowling, Thomas~R}} \yr{2013}  \at{Scaling of
  impact craters in unconsolidated granular materials}.  \jt{American Journal
  of Physics}  \bvol{81}~(11),  \pg{875--878}.

\bibitem[Driscoll \& Nagel(2011)]{driscoll2011ultrafast}
{\sc \au{Driscoll, Michelle~M} \& \au{Nagel, Sidney~R}} \yr{2011}
  \at{Ultrafast interference imaging of air in splashing dynamics}.
  \jt{Physical review letters}  \bvol{107}~(15),  \pg{154502}.

\bibitem[Gilet \& Bourouiba(2015)]{gilet2015fluid}
{\sc \au{Gilet, Tristan} \& \au{Bourouiba, Lydia}} \yr{2015}  \at{Fluid
  fragmentation shapes rain-induced foliar disease transmission}.  \jt{Journal
  of the Royal Society Interface}  \bvol{12}~(104),  \pg{20141092}.

\bibitem[Gillen \& Guha(2005)]{gillen2005use}
{\sc \au{Gillen, Glen~D} \& \au{Guha, Shekhar}} \yr{2005}  \at{Use of michelson
  and fabry--perot interferometry for independent determination of the
  refractive index and physical thickness of wafers}.  \jt{Applied optics}
  \bvol{44}~(3),  \pg{344--347}.

\bibitem[de~Goede {\em et~al.\/}(2019)de~Goede, de~Bruin, Shahidzadeh \&
  Bonn]{de2019predicting}
{\sc \au{de~Goede, Thijs~C}, \au{de~Bruin, Karla~G}, \au{Shahidzadeh, Noushine}
  \& \au{Bonn, Daniel}} \yr{2019}  \at{Predicting the maximum spreading of a
  liquid drop impacting on a solid surface: Effect of surface tension and
  entrapped air layer}.  \jt{Physical review fluids}  \bvol{4}~(5),
  \pg{053602}.

\bibitem[Hariharan {\em et~al.\/}(2022)Hariharan, Chowdhury, Rao~S,
  Chakravortty \& Basu]{Hariharan2022.05.28.493826}
{\sc \au{Hariharan, Vishnu}, \au{Chowdhury, Atish~Roy}, \au{Rao~S, Srinivas},
  \au{Chakravortty, Dipshikha} \& \au{Basu, Saptarshi}} \yr{2022}
  \at{Physiological characteristics of bacterial droplets indicate a
  catastrophic consequence with an increase in impact velocity}.  \jt{bioRxiv}
  ,  \arxiv{arXiv:
  https://www.biorxiv.org/content/early/2022/05/29/2022.05.28.493826.full.pdf}.

\bibitem[Hecht(2012)]{hecht2012optics}
{\sc \au{Hecht, Eugene}} \yr{2012} {\em Optics\/}.  \publ{Pearson Education
  India}.

\bibitem[Hicks \& Purvis(2010)]{hicks2010air}
{\sc \au{Hicks, Peter~D} \& \au{Purvis, Richard}} \yr{2010}  \at{Air cushioning
  and bubble entrapment in three-dimensional droplet impacts}.  \jt{Journal of
  Fluid Mechanics}  \bvol{649},  \pg{135--163}.

\bibitem[Hicks \& Purvis(2011)]{hicks2011air}
{\sc \au{Hicks, Peter~D} \& \au{Purvis, Richard}} \yr{2011}  \at{Air cushioning
  in droplet impacts with liquid layers and other droplets}.  \jt{Physics of
  fluids}  \bvol{23}~(6),  \pg{062104}.

\bibitem[Hocking(1973)]{hocking1973effect}
{\sc \au{Hocking, Leslie~M}} \yr{1973}  \at{The effect of slip on the motion of
  a sphere close to a wall and of two adjacent spheres}.  \jt{Journal of
  Engineering Mathematics}  \bvol{7}~(3),  \pg{207--221}.

\bibitem[How {\em et~al.\/}(2021)How, Koch \& Collins]{how2021non}
{\sc \au{How, Melanie Li~Sing}, \au{Koch, Donald~L} \& \au{Collins, Lance~R}}
  \yr{2021}  \at{Non-continuum tangential lubrication gas flow between two
  spheres}.  \jt{Journal of Fluid Mechanics}  \bvol{920},  \pg{A2}.

\bibitem[Josserand \& Thoroddsen(2016)]{josserand2016drop}
{\sc \au{Josserand, Christophe} \& \au{Thoroddsen, Sigurdur~T}} \yr{2016}
  \at{Drop impact on a solid surface}.  \jt{Annual review of fluid mechanics}
  \bvol{48},  \pg{365--391}.

\bibitem[Kai \& Kemao(2010)]{kai2010fast}
{\sc \au{Kai, Li} \& \au{Kemao, Qian}} \yr{2010}  \at{Fast frequency-guided
  sequential demodulation of a single fringe pattern}.  \jt{Optics letters}
  \bvol{35}~(22),  \pg{3718--3720}.

\bibitem[Kemao \& Soon(2007)]{kemao2007sequential}
{\sc \au{Kemao, Qian} \& \au{Soon, Seah~Hock}} \yr{2007}  \at{Sequential
  demodulation of a single fringe pattern guided by local frequencies}.
  \jt{Optics letters}  \bvol{32}~(2),  \pg{127--129}.

\bibitem[Kim {\em et~al.\/}(2017)Kim, Kim, Kang, Jin \&
  Lee]{kim2017interferometric}
{\sc \au{Kim, Jong-Ahn}, \au{Kim, Jae~Wan}, \au{Kang, Chu-Shik}, \au{Jin,
  Jonghan} \& \au{Lee, Jae~Yong}} \yr{2017}  \at{An interferometric system for
  measuring thickness of parallel glass plates without 2$\pi$ ambiguity using
  phase analysis of quadrature haidinger fringes}.  \jt{Review of Scientific
  Instruments}  \bvol{88}~(5).

\bibitem[Kitagawa(2013)]{kitagawa2013thin}
{\sc \au{Kitagawa, Katsuichi}} \yr{2013}  \at{Thin-film thickness profile
  measurement by three-wavelength interference color analysis}.  \jt{Applied
  optics}  \bvol{52}~(10),  \pg{1998--2007}.

\bibitem[Koch(1990)]{koch1990kinetic}
{\sc \au{Koch, Donald~L}} \yr{1990}  \at{Kinetic theory for a monodisperse
  gas--solid suspension}.  \jt{Physics of Fluids A: Fluid Dynamics}
  \bvol{2}~(10),  \pg{1711--1723}.

\bibitem[Kolinski {\em et~al.\/}(2012)Kolinski, Rubinstein, Mandre, Brenner,
  Weitz \& Mahadevan]{kolinski2012skating}
{\sc \au{Kolinski, John~M}, \au{Rubinstein, Shmuel~M}, \au{Mandre, Shreyas},
  \au{Brenner, Michael~P}, \au{Weitz, David~A} \& \au{Mahadevan, L}} \yr{2012}
  \at{Skating on a film of air: drops impacting on a surface}.  \jt{Physical
  review letters}  \bvol{108}~(7),  \pg{074503}.

\bibitem[Langley {\em et~al.\/}(2017)Langley, Li \&
  Thoroddsen]{langley2017impact}
{\sc \au{Langley, Kenneth}, \au{Li, Er~Qiang} \& \au{Thoroddsen, Sigurdur~T}}
  \yr{2017}  \at{Impact of ultra-viscous drops: air-film gliding and extreme
  wetting}.  \jt{Journal of Fluid Mechanics}  \bvol{813},  \pg{647--666}.

\bibitem[Langley \& Thoroddsen(2019)]{langley2019gliding}
{\sc \au{Langley, KR} \& \au{Thoroddsen, Sigurdur~T}} \yr{2019}  \at{Gliding on
  a layer of air: impact of a large-viscosity drop on a liquid film}.
  \jt{Journal of Fluid Mechanics}  \bvol{878},  \pg{R2}.

\bibitem[Langmuir \& Blodgett(1946)]{langmuir1946mathematical}
{\sc \au{Langmuir, Irving} \& \au{Blodgett, Katherine}} \yr{1946} {\em A
  mathematical investigation of water droplet trajectories\/}.  \publ{Army Air
  Forces Headquarters, Air Technical Service Command}.

\bibitem[Lesser \& Field(1983)]{lesser1983impact}
{\sc \au{Lesser, Martin~B} \& \au{Field, JE}} \yr{1983}  \at{The impact of
  compressible liquids}.  \jt{Annual review of fluid mechanics}  \bvol{15}~(1),
   \pg{97--122}.

\bibitem[Li {\em et~al.\/}(2015)Li, Vakarelski \& Thoroddsen]{li2015probing}
{\sc \au{Li, EQ}, \au{Vakarelski, Ivan~Uriev} \& \au{Thoroddsen, Sigurdur~T}}
  \yr{2015}  \at{Probing the nanoscale: the first contact of an impacting
  drop}.  \jt{Journal of Fluid Mechanics}  \bvol{785},  \pg{R2}.

\bibitem[Limozin \& Sengupta(2009)]{limozin2009quantitative}
{\sc \au{Limozin, Laurent} \& \au{Sengupta, Kheya}} \yr{2009}  \at{Quantitative
  reflection interference contrast microscopy (ricm) in soft matter and cell
  adhesion}.  \jt{ChemPhysChem}  \bvol{10}~(16),  \pg{2752--2768}.

\bibitem[Lohse(2022)]{lohse2022fundamental}
{\sc \au{Lohse, Detlef}} \yr{2022}  \at{Fundamental fluid dynamics challenges
  in inkjet printing}.  \jt{Annual review of fluid mechanics}  \bvol{54},
  \pg{349--382}.

\bibitem[Mandre \& Brenner(2012)]{mandre2012mechanism}
{\sc \au{Mandre, Shreyas} \& \au{Brenner, Michael~P}} \yr{2012}  \at{The
  mechanism of a splash on a dry solid surface}.  \jt{Journal of Fluid
  Mechanics}  \bvol{690},  \pg{148--172}.

\bibitem[Mandre {\em et~al.\/}(2009)Mandre, Mani \&
  Brenner]{mandre2009precursors}
{\sc \au{Mandre, Shreyas}, \au{Mani, Madhav} \& \au{Brenner, Michael~P}}
  \yr{2009}  \at{Precursors to splashing of liquid droplets on a solid
  surface}.  \jt{Physical review letters}  \bvol{102}~(13),  \pg{134502}.

\bibitem[Park {\em et~al.\/}(2019)Park, Kim, Ahn, Bae \& Jin]{park2019review}
{\sc \au{Park, Jungjae}, \au{Kim, Jong-Ahn}, \au{Ahn, Heulbi}, \au{Bae,
  Jaeseok} \& \au{Jin, Jonghan}} \yr{2019}  \at{A review of thickness
  measurements of thick transparent layers using optical interferometry}.
  \jt{International Journal of Precision Engineering and Manufacturing}
  \bvol{20},  \pg{463--477}.

\bibitem[Qi {\em et~al.\/}(2020)Qi, Wang \& Che]{qi2020air}
{\sc \au{Qi, Haicheng}, \au{Wang, Tianyou} \& \au{Che, Zhizhao}} \yr{2020}
  \at{Air layer during the impact of droplets on heated substrates}.
  \jt{Physical Review E}  \bvol{101}~(4),  \pg{043114}.

\bibitem[Roy \& Basu(2022)]{roy2022thermofluidics}
{\sc \au{Roy, Durbar} \& \au{Basu, Saptarshi}} \yr{2022}  \at{On the
  thermofluidics of a steady laminar jet impacting on a rotating hot plate: An
  ab initio scaling perspective}.  \jt{AIP Advances}  \bvol{12}~(8).

\bibitem[Roy {\em et~al.\/}(2019)Roy, Pandey, Banik, Mukherjee \&
  Basu]{roy2019dynamics}
{\sc \au{Roy, Durbar}, \au{Pandey, Khushboo}, \au{Banik, Meneka},
  \au{Mukherjee, Rabibrata} \& \au{Basu, Saptarshi}} \yr{2019}  \at{Dynamics of
  droplet impingement on bioinspired surface: insights into spreading,
  anomalous stickiness and break-up}.  \jt{Proceedings of the Royal Society A}
  \bvol{475}~(2229),  \pg{20190260}.

\bibitem[Roy {\em et~al.\/}(2022)Roy, Rao \& Basu]{roy2022droplet}
{\sc \au{Roy, Durbar}, \au{Rao, Srinivas~S} \& \au{Basu, Saptarshi}} \yr{2022}
  \at{Droplet impact on immiscible liquid pool: Multi-scale dynamics of
  entrapped air cushion at short timescales}.  \jt{Physics of Fluids}
  \bvol{34}~(5),  \pg{052004}.

\bibitem[Roy {\em et~al.\/}(2021)Roy, Rasheed, Kabi, Roy, Shetty \&
  Basu]{roy2021fluid}
{\sc \au{Roy, Durbar}, \au{Rasheed, Abdur}, \au{Kabi, Prasenjit}, \au{Roy,
  Abhijit~Sinha}, \au{Shetty, Rohit} \& \au{Basu, Saptarshi}} \yr{2021}
  \at{Fluid dynamics of droplet generation from corneal tear film during
  non-contact tonometry in the context of pathogen transmission}.  \jt{Physics
  of Fluids}  \bvol{33}~(9),  \pg{092109}.

\bibitem[Roy {\em et~al.\/}(2023)Roy, Sophia \& Basu]{roy2023mechanics}
{\sc \au{Roy, Durbar}, \au{Sophia, M} \& \au{Basu, Saptarshi}} \yr{2023}
  \at{On the mechanics of droplet surface crater during impact on immiscible
  viscous liquid pool}.  \jt{Journal of Fluid Mechanics}  \bvol{955},
  \pg{A27}.

\bibitem[de~Ruiter {\em et~al.\/}(2015{\natexlab{{\em a\/}}})de~Ruiter, van~den
  Ende \& Mugele]{de2015air-2}
{\sc \au{de~Ruiter, Jolet}, \au{van~den Ende, Dirk} \& \au{Mugele, Frieder}}
  \yr{2015{\natexlab{{\em a\/}}}}  \at{Air cushioning in droplet impact. ii.
  experimental characterization of the air film evolution}.  \jt{Physics of
  fluids}  \bvol{27}~(1),  \pg{012105}.

\bibitem[de~Ruiter {\em et~al.\/}(2015{\natexlab{{\em b\/}}})de~Ruiter, Mugele
  \& van~den Ende]{de2015air-1}
{\sc \au{de~Ruiter, Jolet}, \au{Mugele, Frieder} \& \au{van~den Ende, Dirk}}
  \yr{2015{\natexlab{{\em b\/}}}}  \at{Air cushioning in droplet impact. i.
  dynamics of thin films studied by dual wavelength reflection interference
  microscopy}.  \jt{Physics of fluids}  \bvol{27}~(1),  \pg{012104}.

\bibitem[de~Ruiter {\em et~al.\/}(2012)de~Ruiter, Oh, van~den Ende \&
  Mugele]{de2012dynamics}
{\sc \au{de~Ruiter, Jolet}, \au{Oh, Jung~Min}, \au{van~den Ende, Dirk} \&
  \au{Mugele, Frieder}} \yr{2012}  \at{Dynamics of collapse of air films in
  drop impact}.  \jt{Physical review letters}  \bvol{108}~(7),  \pg{074505}.

\bibitem[Schneider {\em et~al.\/}(2012)Schneider, Rasband \&
  Eliceiri]{schneider2012nih}
{\sc \au{Schneider, Caroline~A}, \au{Rasband, Wayne~S} \& \au{Eliceiri,
  Kevin~W}} \yr{2012}  \at{Nih image to imagej: 25 years of image analysis}.
  \jt{Nature methods}  \bvol{9}~(7),  \pg{671--675}.

\bibitem[Shetty {\em et~al.\/}(2020)Shetty, Balakrishnan, Shroff, Shetty, Kabi,
  Roy, Joseph, Khamar, Basu \& Sinha~Roy]{shetty2020quantitative}
{\sc \au{Shetty, Rohit}, \au{Balakrishnan, Nikhil}, \au{Shroff, Sujani},
  \au{Shetty, Naren}, \au{Kabi, Prasenjit}, \au{Roy, Durbar}, \au{Joseph,
  Sophia~M}, \au{Khamar, Pooja}, \au{Basu, Saptarshi} \& \au{Sinha~Roy,
  Abhijit}} \yr{2020}  \at{Quantitative high-speed assessment of droplet and
  aerosol from an eye after impact with an air-puff amid covid-19 scenario}.
  \jt{Journal of Glaucoma}  \bvol{29}~(11),  \pg{1006--1016}.

\bibitem[Shukla {\em et~al.\/}(2006)Shukla, Udupa, Das \&
  Mantravadi]{shukla2006non}
{\sc \au{Shukla, RP}, \au{Udupa, DV}, \au{Das, NC} \& \au{Mantravadi, Murty~V}}
  \yr{2006}  \at{Non-destructive thickness measurement of dichromated gelatin
  films deposited on glass plates}.  \jt{Optics \& Laser Technology}
  \bvol{38}~(7),  \pg{552--557}.

\bibitem[Sugiyama {\em et~al.\/}(2006)Sugiyama, Ogawa, Kitagawa \&
  Suzuki]{sugiyama2006single}
{\sc \au{Sugiyama, Masashi}, \au{Ogawa, Hidemitsu}, \au{Kitagawa, Katsuichi} \&
  \au{Suzuki, Kazuyoshi}} \yr{2006}  \at{Single-shot surface profiling by local
  model fitting}.  \jt{Applied Optics}  \bvol{45}~(31),  \pg{7999--8005}.

\bibitem[Sundararajakumar \& Koch(1996)]{sundararajakumar1996non}
{\sc \au{Sundararajakumar, RR} \& \au{Koch, Donald~L}} \yr{1996}
  \at{Non-continuum lubrication flows between particles colliding in a gas}.
  \jt{Journal of Fluid Mechanics}  \bvol{313},  \pg{283--308}.

\bibitem[Thoroddsen {\em et~al.\/}(2008)Thoroddsen, Etoh \&
  Takehara]{thoroddsen2008high}
{\sc \au{Thoroddsen, Sigurdur~T}, \au{Etoh, Takeharu~Goji} \& \au{Takehara,
  Kohsei}} \yr{2008}  \at{High-speed imaging of drops and bubbles}.  \jt{Annu.
  Rev. Fluid Mech.}  \bvol{40},  \pg{257--285}.

\bibitem[Thoroddsen {\em et~al.\/}(2012)Thoroddsen, Takehara \&
  Etoh]{thoroddsen2012micro}
{\sc \au{Thoroddsen, Sigurdur~T}, \au{Takehara, Kohsei} \& \au{Etoh, TG}}
  \yr{2012}  \at{Micro-splashing by drop impacts}.  \jt{Journal of Fluid
  Mechanics}  \bvol{706},  \pg{560--570}.

\bibitem[Valentini {\em et~al.\/}(2023)Valentini, Verhoff, Grover \&
  Bisek]{valentini2023first}
{\sc \au{Valentini, Paolo}, \au{Verhoff, Ashley~M}, \au{Grover, Maninder~S} \&
  \au{Bisek, Nicholas~J}} \yr{2023}  \at{First-principles predictions for shear
  viscosity of air components at high temperature}.  \jt{Physical Chemistry
  Chemical Physics}  \bvol{25}~(13),  \pg{9131--9139}.

\bibitem[Van Der~Veen {\em et~al.\/}(2012)Van Der~Veen, Tran, Lohse \&
  Sun]{van2012direct}
{\sc \au{Van Der~Veen, Roeland~CA}, \au{Tran, Tuan}, \au{Lohse, Detlef} \&
  \au{Sun, Chao}} \yr{2012}  \at{Direct measurements of air layer profiles
  under impacting droplets using high-speed color interferometry}.
  \jt{Physical Review E}  \bvol{85}~(2),  \pg{026315}.

\bibitem[Van~Rossum \& Drake(2009)]{10.5555/1593511}
{\sc \au{Van~Rossum, Guido} \& \au{Drake, Fred~L.}} \yr{2009} {\em Python 3
  Reference Manual\/}.  \publ{Scotts Valley, CA: CreateSpace}.

\bibitem[Vrij(1966)]{vrij1966possible}
{\sc \au{Vrij, Agienus}} \yr{1966}  \at{Possible mechanism for the spontaneous
  rupture of thin, free liquid films}.  \jt{Discussions of the Faraday Society}
   \bvol{42},  \pg{23--33}.

\bibitem[Wang \& Kemao(2009)]{wang2009frequency}
{\sc \au{Wang, Haixia} \& \au{Kemao, Qian}} \yr{2009}  \at{Frequency guided
  methods for demodulation of a single fringe pattern}.  \jt{Optics Express}
  \bvol{17}~(17),  \pg{15118--15127}.

\bibitem[Wang \& Bourouiba(2018)]{wang2018non}
{\sc \au{Wang, Y} \& \au{Bourouiba, L}} \yr{2018}  \at{Non-isolated drop impact
  on surfaces}.  \jt{Journal of Fluid Mechanics}  \bvol{835},  \pg{24--44}.

\bibitem[Wijshoff(2018)]{wijshoff2018drop}
{\sc \au{Wijshoff, Herman}} \yr{2018}  \at{Drop dynamics in the inkjet printing
  process}.  \jt{Current opinion in colloid \& interface science}  \bvol{36},
  \pg{20--27}.

\bibitem[Woolf {\em et~al.\/}(2007)Woolf, Leifer, Nightingale, Rhee, Bowyer,
  Caulliez, De~Leeuw, Larsen, Liddicoat, Baker {\em
  et~al.\/}]{woolf2007modelling}
{\sc \au{Woolf, DK}, \au{Leifer, IS}, \au{Nightingale, PD}, \au{Rhee, TS},
  \au{Bowyer, P}, \au{Caulliez, Guillemette}, \au{De~Leeuw, G}, \au{Larsen,
  S{\o}ren~Ejling}, \au{Liddicoat, M}, \au{Baker, J} \& \au{others}} \yr{2007}
  \at{Modelling of bubble-mediated gas transfer: Fundamental principles and a
  laboratory test}.  \jt{Journal of Marine Systems}  \bvol{66}~(1-4),
  \pg{71--91}.

\bibitem[Worthington(1877)]{worthington1877xxviii}
{\sc \au{Worthington, Arthur~Mason}} \yr{1877}  \at{Xxviii. on the forms
  assumed by drops of liquids falling vertically on a horizontal plate}.
  \jt{Proceedings of the royal society of London}  \bvol{25}~(171-178),
  \pg{261--272}.

\bibitem[Worthington(1883)]{worthington1883impact}
{\sc \au{Worthington, Arthur~Mason}} \yr{1883}  \at{On impact with a liquid
  surface}.  \jt{Proceedings of the Royal Society of London}
  \bvol{34}~(220-223),  \pg{217--230}.

\bibitem[Worthington \& Cole(1897)]{worthington1897v}
{\sc \au{Worthington, Arthur~Mason} \& \au{Cole, Reginald~Sorr{\`e}}} \yr{1897}
   \at{V. impact with a liquid surface, studied by the aid of instantaneous
  photography}.  \jt{Philosophical Transactions of the Royal Society of London.
  Series A, Containing Papers of a Mathematical or Physical Character} ~(189),
  \pg{137--148}.

\bibitem[Worthington \& Cole(1900)]{worthington1900iv}
{\sc \au{Worthington, Arthur~Mason} \& \au{Cole, Reginald~Sorr{\`e}}} \yr{1900}
   \at{Iv. impact with a liquid surface studied by the aid of instantaneous
  photography. paper ii}.  \jt{Philosophical Transactions of the Royal Society
  of London. Series A, Containing Papers of a Mathematical or Physical
  Character}  \bvol{194}~(252-261),  \pg{175--199}.

\bibitem[Worthington {\em et~al.\/}(1877)]{worthington1877second}
{\sc \au{Worthington, Arthur~M} \& \au{others}} \yr{1877}  \at{A second paper
  on the forms assumed by drops of liquids falling vertically on a horizontal
  plate}.  \jt{Proc. R. Soc. London}  \bvol{25},  \pg{498--503}.

\bibitem[Xu {\em et~al.\/}(2005)Xu, Zhang \& Nagel]{xu2005drop}
{\sc \au{Xu, Lei}, \au{Zhang, Wendy~W} \& \au{Nagel, Sidney~R}} \yr{2005}
  \at{Drop splashing on a dry smooth surface}.  \jt{Physical review letters}
  \bvol{94}~(18),  \pg{184505}.

\bibitem[Yarin(2006)]{yarin2006drop}
{\sc \au{Yarin, Alexander~L}} \yr{2006}  \at{Drop impact dynamics: splashing,
  spreading, receding, bouncing…}.  \jt{Annu. Rev. Fluid Mech.}  \bvol{38},
  \pg{159--192}.

\bibitem[Yarin {\em et~al.\/}(2017)Yarin, Roisman \&
  Tropea]{yarin2017collision}
{\sc \au{Yarin, Alexander~L}, \au{Roisman, Ilia~V} \& \au{Tropea, Cameron}}
  \yr{2017} {\em Collision phenomena in liquids and solids\/}.  \publ{Cambridge
  University Press}.

\end{thebibliography}

\end{document}